\documentclass[lettersize,journal]{IEEEtran}
\usepackage{amsmath,amsfonts}
\usepackage{algorithmic}
\usepackage{algorithm}
\usepackage{array}
\usepackage[caption=false,font=normalsize,labelfont=sf,textfont=sf]{subfig}
\usepackage{textcomp}
\usepackage{stfloats}
\usepackage{url}
\usepackage{verbatim}
\usepackage{graphicx}
\usepackage{cite}
\usepackage{tabularx}
\usepackage[table]{xcolor}
\usepackage{amssymb}
\usepackage{amsmath,amssymb,amsfonts}
\usepackage{comment}
\usepackage{multirow}
\usepackage{makecell}
\usepackage[table,xcdraw]{xcolor} 
\usepackage{array}
\usepackage{lscape} 
\definecolor{darkgreen}{RGB}{0,128,0} 
\usepackage{bm}

\begin{document}

\title{Large Language Models and Attention-Based AI for Hardware Design and Security: Progress, Challenges, and Opportunities}

\author{
    Sujan Ghimire\textsuperscript{1}, 
    Parsa Mirfasihi \textsuperscript{1},
    Muhtasim Alam Chowdhury\textsuperscript{1},
    Harish Kumar Dharavath\textsuperscript{1} ,\\
    Banafsheh Saber Latibari\textsuperscript{1}, 
    Muntasir Mamun\textsuperscript{1},
    Jaeden Wolf Carpenter\textsuperscript{1}, 
    Benjamin Tan\textsuperscript{3}, 
    Hammond Pearce\textsuperscript{4}, 
    Farshad Firouzi\textsuperscript{5},
    Krishnendu Chakrabarty\textsuperscript{5},
    Rozhin Yasaei \textsuperscript{6},
    Pratik Satam\textsuperscript{1,2}, 
    and Soheil Salehi\textsuperscript{1}
    \thanks{\textsuperscript{1}Electrical and Computer Engineering Department, The University of Arizona, Tucson, AZ, USA (e-mail:sghimire@arizona.edu, mmc7@arizona.edu, banafsheh@arizona.edu,  muntasir@arizona.edu, carpenterjaeden@arizona.edu, ssalehi@arizona.edu).}
    
    \thanks{\textsuperscript{2}Systems and Industrial Engineering Department, The University of Arizona, Tucson, AZ, USA (e-mail: pratiksatam@arizona.edu).}%
    \thanks{\textsuperscript{3}Department of Electrical and Software Engineering, University of Calgary, Calgary, Alberta, CA (e-mail: benjamin.tan1@ucalgary.ca).}%
    \thanks{\textsuperscript{4}School of Computer Science and Engineering, University of New South Wales, Sydney, AU (e-mail: hammond.pearce@unsw.edu.au).}%
    \thanks{\textsuperscript{5}School of Electrical Computer and Energy Engineering, Arizona State University, Tempe, AZ, USA (e-mail: Farshad.Firouzi@asu.edu, Krishnendu.Chakrabarty@asu.edu).}
    \thanks{\textsuperscript{6}College of Information Science, The University of Arizona, Tucson, AZ 85721 USA (e-mail: yasaei@arizona.edu).}

   }

\markboth{}%
{Shell \MakeLowercase{\textit{et al.}}: A Sample Article Using IEEEtran.cls for IEEE Journals}


\maketitle

\begin{abstract}
Recent advances in attention-based artificial intelligence (AI) models have unlocked vast potential to automate digital hardware design while enhancing and strengthening security measures against various threats. This rapidly emerging field leverages Large Language Models (LLMs) to generate HDL code, identify vulnerabilities, and sometimes mitigate them. The state of the art in this design automation space utilizes optimized LLMs with HDL datasets, creating automated systems for register transfer level (RTL) generation, verification, and debugging, and establishing LLM-driven design environments for streamlined logic designs. Additionally, attention-based models like graph attention have shown promise in chip design applications, including floorplanning. This survey investigates the integration of these models into hardware-related domains, emphasizing logic design and hardware security, with or without the use of IP libraries. This study explores the commercial and academic landscape, highlighting technical hurdles and future prospects for automating hardware design and security. 
Moreover, it provides new insights into the study of LLM-driven design systems, advances in hardware security mechanisms, and the impact of influential works on industry practices. Through the examination of 30 representative approaches and illustrative case studies, this paper underscores the transformative potential of attention-based models in revolutionizing hardware design while addressing the challenges that lie ahead in this interdisciplinary domain. 
\end{abstract}


\section{Introduction}\label{sec:intro}
\IEEEPARstart{H}{ardware} design is a complex process that begins with specific requirements outlined in natural language. It involves hardware engineers translating these requirements into Hardware Description Languages (HDLs) before verifying them extensively~\cite{yang2023new,tomlinson2024designing}. During this process, hardware engineers must ensure that design constraints (such as performance, power, and area) are met, which can be repetitive and prone to errors and inefficiencies. 
Automating these processes with machine learning (ML) models, such as recent deep neural networks featuring the \textit{attention} mechanism~\cite{NIPS2017_3f5ee243}, can significantly reduce the likelihood of human error and accelerate the design cycle by streamlining tasks and improving design efficiency. Initially, the attention-based model was introduced as an alternative to recurrent neural networks (RNNs) for natural language processing (NLP)~\cite{NIPS2017_3f5ee243}. 
The transformer model, which uses self-attention to focus on different parts of input data to make predictions, has led to pioneering outcomes in many domains like speech recognition, computer vision, sentiment analysis, named entity recognition, and time series analysis~\cite{10.1145/3649476.3660380, app11093883, latibari2024transformers}. LLMs such as the Generative Pre-trained Transformer (GPT) series have demonstrated remarkable capabilities in generating human-like text, understanding context, and performing complex language-related tasks~\cite{openai2024gpt4}.

\begin{figure}[t]
    \centering
    \includegraphics[width=1\linewidth]{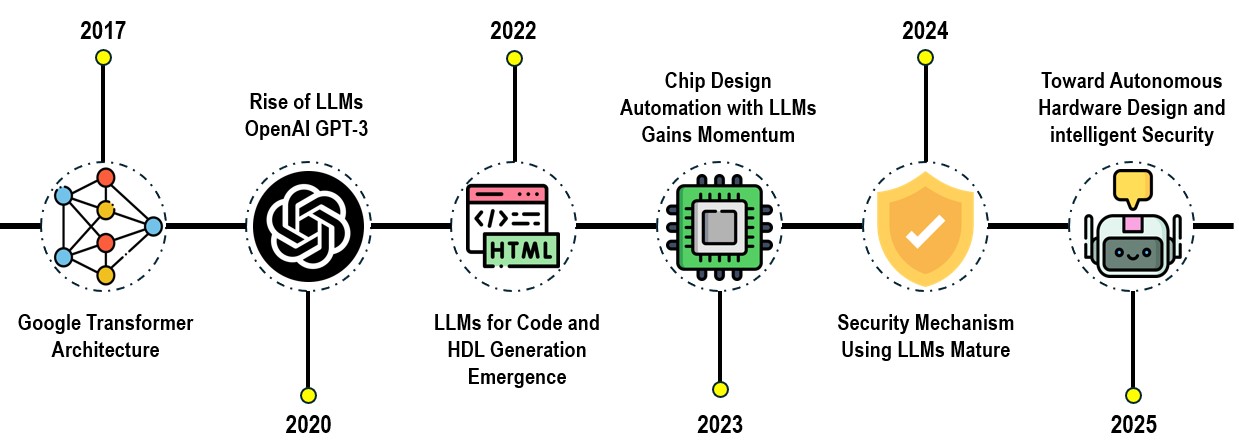}
    \caption{Timeline of AI advancements and their integration into hardware design and security from 2017 to 2025. } 
    \label{fig:timeline}
\end{figure}

Figure~\ref{fig:timeline} shows the evolution of key advancements in AI and hardware design, from the introduction of the Google Transformer architecture (2017) to the rise of LLMs like GPT-3 (2020), their application in chip design (2023), the first hardware tapeout of hardware designed with LLMs (2023), and advancements in security mechanisms using LLMs (2024) and the deployment of autonomous zero-day discovery agents (2025-2026).

\begin{figure}[!b]
    \centering
    \includegraphics[width=0.95\linewidth]{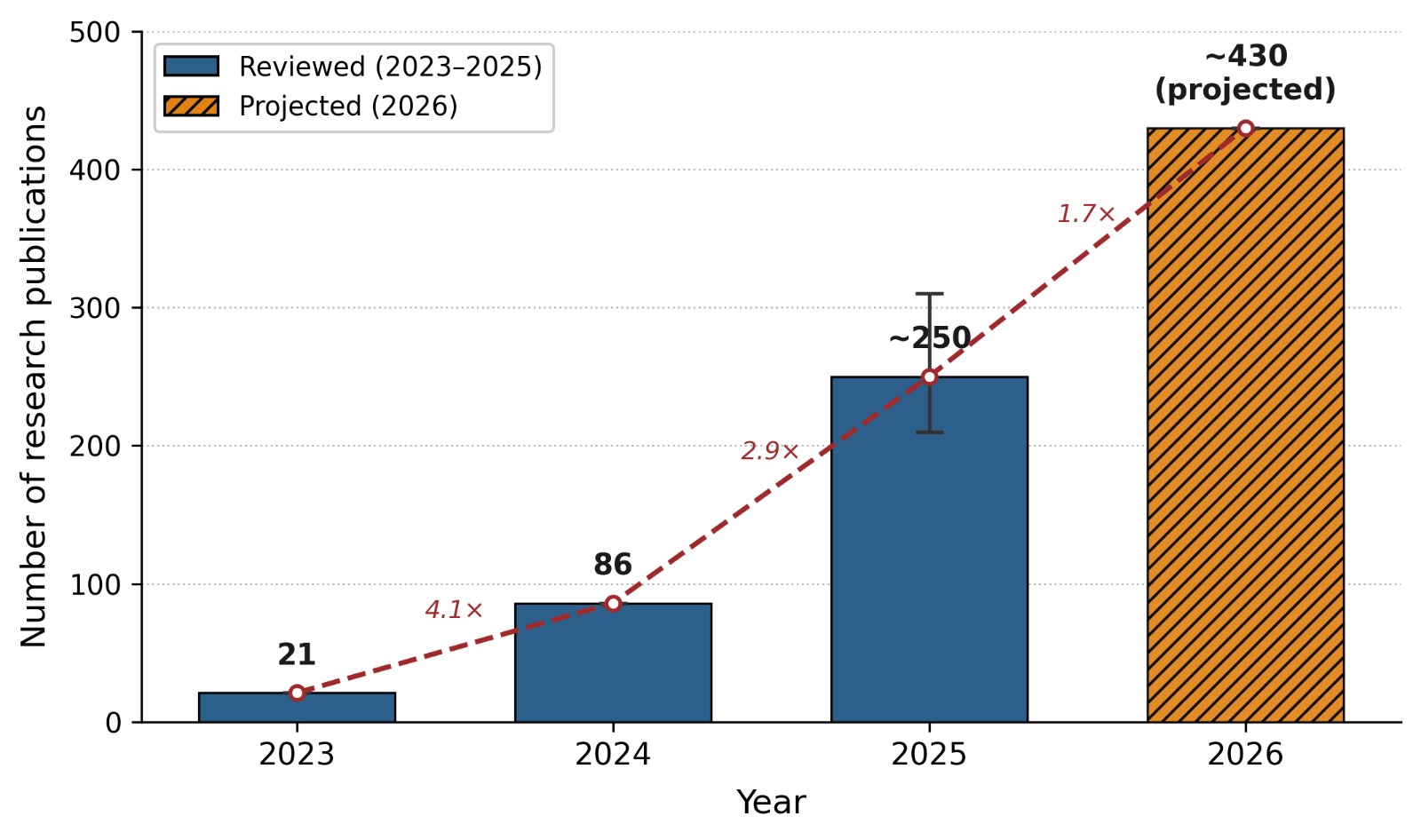}
    \caption{Updated growth trend in research publications on LLM-based and attention-based hardware design and hardware security from 2023 to 2026, based on the reviewed 2025 literature and a projected 2026 continuation.
    } 
    \label{fig:projection2025}
\end{figure}

In the realm of hardware design and security, attention-based models present a transformative opportunity that remains in the early stages of exploration. Figure~\ref{fig:projection2025} illustrates the publication trend in this area from 2023 to 2025, along with a projection for 2026. The reported values are derived from a comprehensive review of the literature surveyed in this paper, which covers studies applying attention-based models to hardware design and security. As shown in the figure, the number of publications increased nearly fourfold, from 21 in 2023 to 86 in 2024. Based on our review of the 2025 literature, this rapid growth continued but began to decelerate, reaching approximately 250 publications in 2025 rather than the roughly 345 publications that would result from a naive continuation of the previous fourfold growth rate. Extrapolating this decelerating trend, we project that the number of publications will reach approximately 430 by the end of 2026, indicating sustained and growing interest in this interdisciplinary research area.

Recent work has explored the potential of LLMs for automated HDL generation from instructions in Natural Language (known as `prompts')~\cite{10299874}. Although this showed potential for speeding up the design process (including successful tape-out) by prompting LLMs, significant human intervention was still required, as pre-trained attention-based models could not handle all the tasks necessary for a complete tape-out design without errors. 

Overcoming this performance gap between LLMs and humans is not trivial. 
One of the significant challenges here is the relative scarcity of public HDL code bases usable for LLM fine-tuning. 
Notable initiatives include optimizing previously pretrained LLMs using Verilog datasets, creating automated frameworks such as Autochip~\cite{thakur2023autochip} for bug correction in LLM-generated HDL code, and creating LLM-based design environments. These initiatives represent initial steps in adapting LLMs to the hardware engineering landscape, focusing on logic design optimization through natural language interaction~\cite{thakur2023autochip}. Other research using attention-based models for the rest of the chip design stages has also faced the challenge of limited data. 

\begin{figure*}[!t]
    \centering
   \includegraphics[trim=1cm 2cm 1cm 2cm, clip, width=0.9\textwidth]{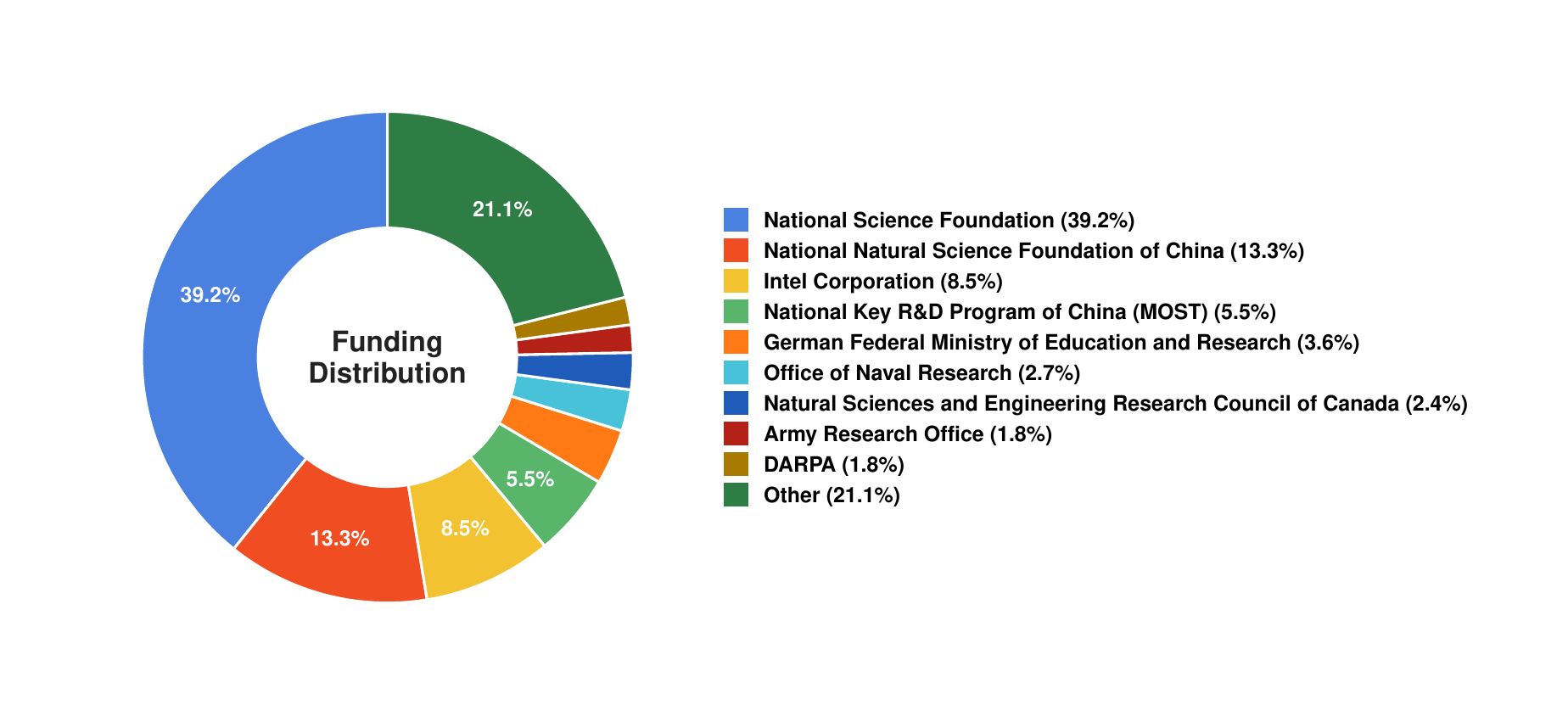}
    \caption{Distribution of research funding across major organizations supporting hardware design, verification, security, and optimization. The chart illustrates the relative contributions of key funding agencies, highlighting the diversity of support across national, governmental, and industrial sources.} 
    \label{fig:funding}
\end{figure*}
 
To address the aforementioned limitations and make the semiconductor industry more efficient, significant interest and support have been directed toward this research area. Figure \ref{fig:funding} shows a snapshot of some of the funding resources supporting this research. To compile this data, we targeted the references that are cited in this paper to identify the key funding agencies supporting research in attention-based hardware design and security. The data reflects a total of  38 funding instances across various research areas, with notable contributions from leading organizations such as NSF, NSFC, SRC, and DARPA. Furthermore, the funding distribution underscores the growing research interest in attention-based hardware design and security, reflecting its increasing significance in shaping future advancements in the field.

In recent work, Bei Yu summarized the 
application of ML techniques in Electronic Design Automation (EDA)~\cite{yu2023machine}. 
He highlighted their impact on various stages of chip design and proposed an adaptive integrated ML paradigm to address challenges such as netlist representation, timing modeling, and multimodal fusion. 
LLM4EDA provided a systematic study on the application of LLMs in EDA, exploring their potential to automate various stages of chip design and improve power, performance, and area (PPA) metrics while highlighting future research directions and maintaining an open-source repository of relevant papers and datasets~\cite{zhong2023llm4eda}. 
Tsung-Yi Ho et al. explored the development and application of AI-native large circuit models (LCMs) in EDA, highlighting their potential to revolutionize the field through comprehensive analysis and creation capabilities while addressing challenges such as data scarcity and scalability~\cite{chen2024dawn}. 
Sharma et al. conducted a thorough analysis of the total cost of ownership (TCO) and performance between ChipNeMo, a domain-adaptive LLM, and leading general-purpose LLMs (Claude 3 Opus and ChatGPT-4 Turbo), demonstrating significant cost savings and performance benefits for ChipNeMo in chip design coding tasks~\cite{10691849}. 
Nan Wu et al. provided a comprehensive overview of ML methodologies applied to hardware design and verification, identifying challenges and proposing solutions within EDA, highlighting the potential and limitations of ML in enhancing design efficiency and accuracy~\cite{10.1145/3661308}. 
Yang et al. presented an in-depth review of LLMs, focusing on their applications, capabilities, challenges, and prospects~\cite{yang2023harnessing}. It explores evaluation metrics like BLEU scores and human-aligned evaluations, emphasizes the importance of prompt engineering for LLM efficacy, compares fine-tuning with prompt tuning, discusses scalability and generalization capabilities, and addresses ethical concerns such as bias and misinformation.

An emerging challenge in hardware design is the need to incorporate security. 
Akyash et al. explored LLMs in hardware security, detailing their use for RTL vulnerability detection and mitigation, evaluation frameworks, comparative analyses, case studies, and future needs for specialized architectures, improved datasets, and EDA tool integration~\cite{akyash2024evolutionary}. 
Similarly, Pearce et al. argue for the potential of LLMs in this space, highlighting successes already obtained by LLMs working in limited hardware security tasks and scenarios~\cite{pearce2024large}.

This paper offers a complementary extension to prior surveys. In considering all types of models based on attention mechanisms, we do not limit our survey purely to LLMs, and we 
survey the use of these models in all design stages, from RTL generation and verification to layout design and routing. 
Further, we provide insights into both hardware \textit{design} and hardware \textit{security}, such as Trojan insertion and logic obfuscation. 


This survey is motivated by a gap that is not fully addressed by prior reviews focusing separately on LLM-based hardware design or LLM-based hardware security. Existing surveys provide valuable overviews of design automation, EDA-oriented LLM systems, or hardware-security applications in isolation, but they provide limited discussion of how attention-based methods operate across the full hardware lifecycle, from specification and RTL generation to verification, physical design, and security analysis. They also insufficiently examine the convergence of hardware design and hardware security when security is treated as a first-class design objective rather than a downstream verification task. Furthermore, existing surveys also do not systematically distinguish between mature capabilities, current limitations, and open research gaps across attention-based hardware design and hardware-security workflows. By surveying both design and security under a common attention-based perspective, this paper aims to answer these cross-cutting questions and provide a more unified view of how LLMs, transformers, and graph-attention models are reshaping hardware workflows~\cite{akyash2024evolutionary,chen2024dawn,pearce2024large,yu2023machine,zhong2023llm4eda}.

\begin{figure}     
    \centering
    \includegraphics[width=1\linewidth]{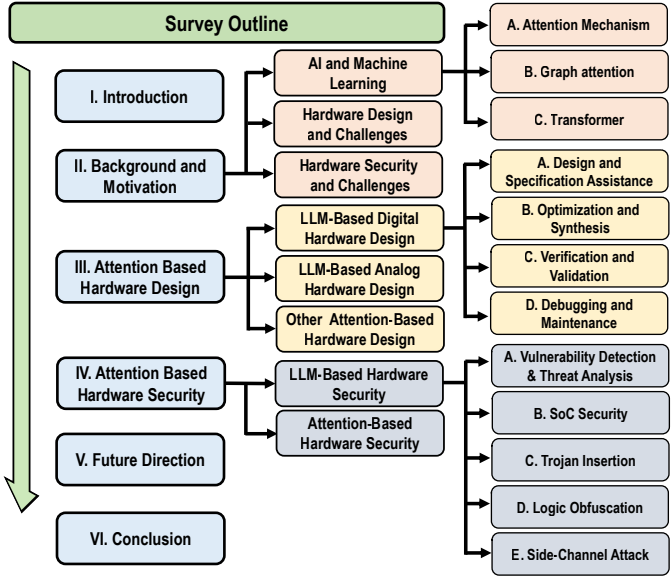}
    \caption{Overview of the survey in this manuscript.}
    \label{overview_dig}
\end{figure}

This survey is intended as a scoped and representative review rather than an exhaustive systematic review of all AI-for-hardware literature. Our primary focus is on recent work that applies large language models, transformers, graph attention models, or other attention-based architectures directly to hardware design, electronic design automation, or hardware security problems. We include papers that make a substantive technical contribution through a method, benchmark, dataset, case study, or integrated design framework, and we organize them according to where they act in the hardware workflow. We exclude general AI papers without a direct hardware-design or hardware-security focus, works in which attention is only peripheral rather than central to the method, and purely software-security studies that do not engage hardware artifacts. Foundational pre-2023 papers are cited selectively when needed for background or historical context, but the main survey emphasizes recent work reflecting the rapid growth of attention-based methods in hardware design and security.

Figure \ref{overview_dig} shows the organization of this paper. 
Section \ref{sec:background} provides background on ML models used in the literature and discusses the limitations and challenges that necessitate the use of these attention-based models in hardware design and security. Sections \ref{sec:AttentionBasedHardwareDesign} and \ref{sec:AttentionBasedHardwareSecurity} review attention-based approaches for hardware design and security, respectively. Future research directions are discussed in Section \ref{sec:discussion}, and Section \ref{sec:conclusion} concludes the paper.

\section{Background and Motivation}
\label{sec:background}
In this section, we provide a comprehensive background on the evolution of AI and  ML, with a particular focus on the transformative advancements in neural network architectures. We begin by introducing the foundational concepts of AI and ML, then discuss graph attention mechanisms and the revolutionary transformer model, highlighting their development and impact over the years. Furthermore, we offer an in-depth overview of these concepts, emphasizing their significance in modern computational paradigms. Subsequently, we transition to the domain of hardware design and explore the challenges it entails. Finally, we address the critical aspect of hardware security, examining the unique vulnerabilities and challenges in hardware design.

\subsection{AI and Machine Learning}

\begin{figure*}[h]     
    \centering
    \includegraphics[width=\linewidth]{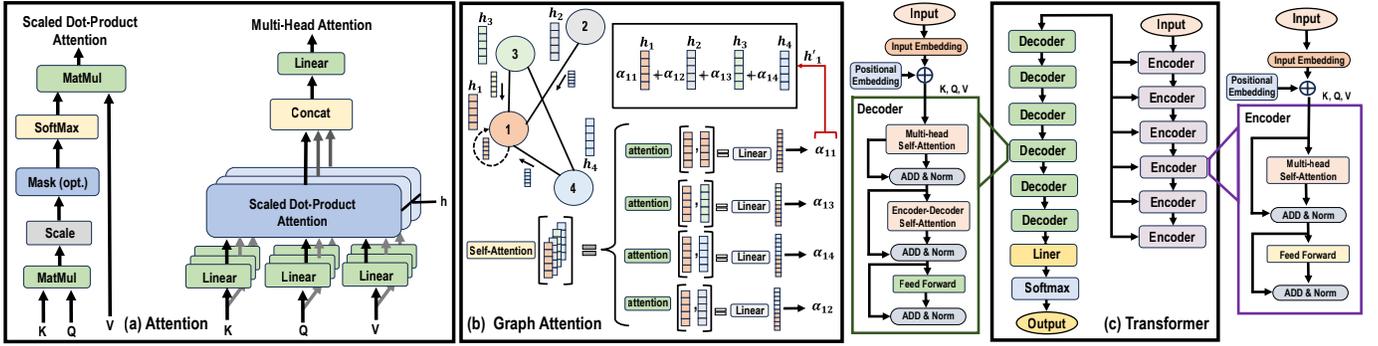}
    \caption{Illustration of the core mechanisms of modern deep learning frameworks: (a) Attention, including scaled dot-product and multi-head attention, (b) Graph Attention for processing graph-based data, and (c) Transformer architecture with encoder-decoder structure.}
    \label{fig:transformer}
\end{figure*}

\subsubsection{Attention Mechanism} 
The attention mechanism is a neural network structure engineered to concentrate on the most relevant aspects of the input data during prediction. 
Bahdanau et al. (2015) initially developed it in neural machine translation to overcome the shortcomings of conventional models such as Recurrent Neural Networks (RNNs), which struggle to capture long-range relationships and manage variable-length sequences~\cite{38ed090f8de9}. 
The core principle of attention is computing a weighted sum of input items, where the weights reflect the importance of each input component. 
This enables the model to focus on critical elements of the input, enhancing its ability to comprehend complex data.  Various types of attention processes are available, such as global attention, which considers the whole input sequence, and local attention, which concentrates on certain portions of the sequence. Self-attention enables each element in the input to focus on all other elements, constituting a fundamental aspect of transformer models. 
Furthermore, multi-head attention utilizes many attention processes concurrently, allowing the model to discern several facets of the input. 
Figure \ref{fig:transformer}-(a) illustrates the attention mechanism focusing on the scaled dot-product attention. The mechanism begins with the inputs, queries ($Q$), keys ($K$), and values ($V$). Then, it requires computing the similarity in scores between the query and keys via matrix multiplication ($QK^T$), which is then scaled by the square root of the key dimensions to stabilize gradients during training. Masking can be applied to selectively ignore specific parts of the input, such as future tokens, in sequence prediction tasks. To produce attention weights, the similarity scores are normalized by using the softmax function. This also highlights the importance of each key relative to the query. The Values ($V$) are then combined with the weighted sum, which is guided by these attention weights. The Multi-head attention model works like an extension by running multiple attention computations in parallel, capturing diverse contextual relationships within the system. 

\subsubsection{Graph Attention}
Graph Attention Networks (GATs) are a neural network architecture specifically engineered for graph-structured data, employing the attention mechanism to overcome the shortcomings of previous graph neural networks (GNNs)~\cite{veličković2018graphattentionnetworks}. GATs utilize self-attention layers rather than fixed convolution procedures, enabling each node to concentrate on its most pertinent neighbors selectively. This is accomplished by calculating attention scores that signify the significance of adjacent nodes according to their attributes. The model subsequently consolidates these attributes, weighted by the attention ratings, to enhance the node's representation. 
GATs do not require the complete graph structure in advance and may accommodate variable-sized neighborhoods, rendering them quite adaptable. Moreover, multi-head attention captures many dimensions of node interactions, augmenting the network's ability to understand intricate patterns. 
GATs have demonstrated robust efficacy in node categorization, connection prediction, and community discovery across several domains, including social networks, biological networks, and citation graphs.
Furthermore, GATs are useful for assessing circuit interdependence and finding possible flaws. By conceptualizing circuit parts (e.g., gates, flip-flops) as nodes and their interactions as edges, GATs can model whole circuits. 

Figure \ref{fig:transformer}-(b) illustrates the overview of the graph attention mechanism, which is the heart of GATs. The figure showcases how each node computes attention scores to focus on its most relevant neighbors by employing this attention mechanism. It does this specifically by computing attention coefficients ($\alpha$), which signify the importance of each neighboring node based on features ($h_1,h_2,h_3,...$). These coefficients are also used in generating an updated representation ($h'_1$) for the target node. As shown in this figure, \ref{fig:transformer}-(b), the process also involves the multiple attention heads that compute various perspectives of node interactions. This approach enhances the model's ability by identifying the complex patterns in graph-structured data.

\subsubsection{Transformer}

The transformer is a neural network architecture that uses self-attention mechanisms, eliminating traditional recurrent and convolutional layers. It consists of an encoder and a decoder, each built with layers of self-attention and feed-forward networks. The self-attention mechanism allows the model to capture relationships across entire input sequences in a highly parallelizable manner, improving computational efficiency. Multi-head attention enables the model to focus on multiple input segments simultaneously, enhancing tasks like machine translation and text generation. The encoder-decoder transformer is ideal for sequence-to-sequence applications, such as machine translation. The encoder in this architecture converts the input sequence into a context vector, which the decoder uses to generate the output sequence. Encoder-based transformers, such as BERT~\cite{devlin-etal-2019-bert}, are designed for tasks requiring an understanding of entire input sequences, like text classification and entity recognition. Decoder-based transformers, such as GPT, specialize in sequence generation, predicting output tokens sequentially in an autoregressive manner by analyzing relationships among previously generated tokens.

The architecture, illustrated in Figure \ref{fig:transformer}-(c), consists of encoder and decoder modules with self-attention and feed-forward layers. The encoder captures contextual relationships within input sequences, while the decoder generates output sequences in an autoregressive manner. Transformers architecture also supports hardware design and security by enabling efficient sequence modeling, dependency extraction, circuit representation learning, Verilog code generation, vulnerability identification, and mitigation generation in hardware designs.

\subsection{Motivation for Attention-Based Methods}

While attention-based models originated in natural language processing, their relevance to hardware design and security arises from structural characteristics of these domains rather than from language alone. Hardware artifacts exhibit long-range dependencies across modules, control paths, interfaces, and timing constraints, and they are represented at multiple abstraction levels, including natural-language specifications, RTL, netlists, graphs, and physical layouts. In digital design, the correctness of a module often depends on signals and constraints defined elsewhere in the architecture or codebase. Hardware security analysis similarly requires correlating dispersed information such as specification intent, privilege boundaries, interface interactions, and side effects that emerge across abstraction layers. Self-attention is therefore attractive because it enables selective reasoning over long contexts, whereas graph-attention mechanisms are well-suited to circuit and netlist representations in which the importance of a node depends strongly on its structural neighborhood. Nevertheless, attention-based approaches should not be viewed as universally superior to lightweight task-specific methods. Their primary advantage emerges in tasks requiring cross-abstraction analysis, natural-language-guided design assistance, graph-structured circuit reasoning, or coordinated processing across multiple stages of the hardware design and security workflow~\cite{delorenzo2024creativeval,NIPS2017_3f5ee243,veličković2018graphattentionnetworks,zhong2023llm4eda}.

\subsection{Hardware Design and Challenges }

\begin{figure*}[t]     
    \centering
    \includegraphics[width=0.95\linewidth]{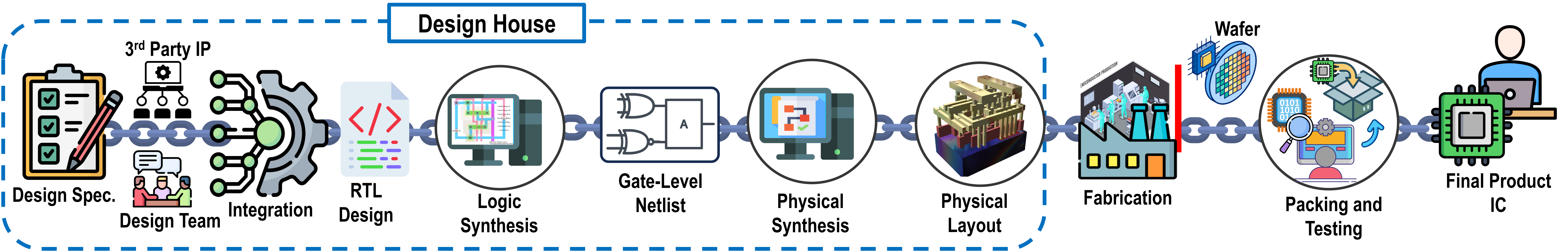}
    \caption{Overview of the key stages in the globalized IC design flow. The Design House progresses from Design Specification to RTL Integration, followed by Logic and Physical Synthesis, and the Final layout. Which transitions to the physical implementation stages of Fabrication, Packing and Testing.}
    \label{fig:hardwaredesign}
\end{figure*}

The hardware design flow is a systematic process that converts high-level design specifications into a functioning integrated circuit ready for implementation, as shown in Figure \ref{fig:hardwaredesign}. The process starts by delineating functional and performance criteria in the design specification~\cite{yang2023new,tomlinson2024designing}. The design team incorporates third-party intellectual property (IP) blocks to accelerate development. The design is captured at the RTL using HDLs such as Verilog or VHDL, emphasizing functionality. The RTL description undergoes synthesis into a gate-level netlist through logic synthesis, which translates the design into fundamental logic gates while optimizing for performance, power, and area. This netlist undergoes physical synthesis, encompassing placement and routing, to produce a comprehensive physical layout. The design is manufactured onto silicon wafers using sophisticated semiconductor techniques, such as photolithography, ion implantation, and chemical vapor deposition, to precisely pattern and build the intricate layers of the integrated circuit. Upon fabrication, the wafers are encased for protection and seamless integration, followed by stringent testing to verify compliance with functional and performance standards. The design team delivers the evaluated integrated circuits as the final product for incorporation into electronic systems. This rigorous process ensures the attainment of design objectives while minimizing mistakes and inefficiencies.

\begin{figure}[!htbp]
    \centering
    \includegraphics[width=0.95\linewidth]{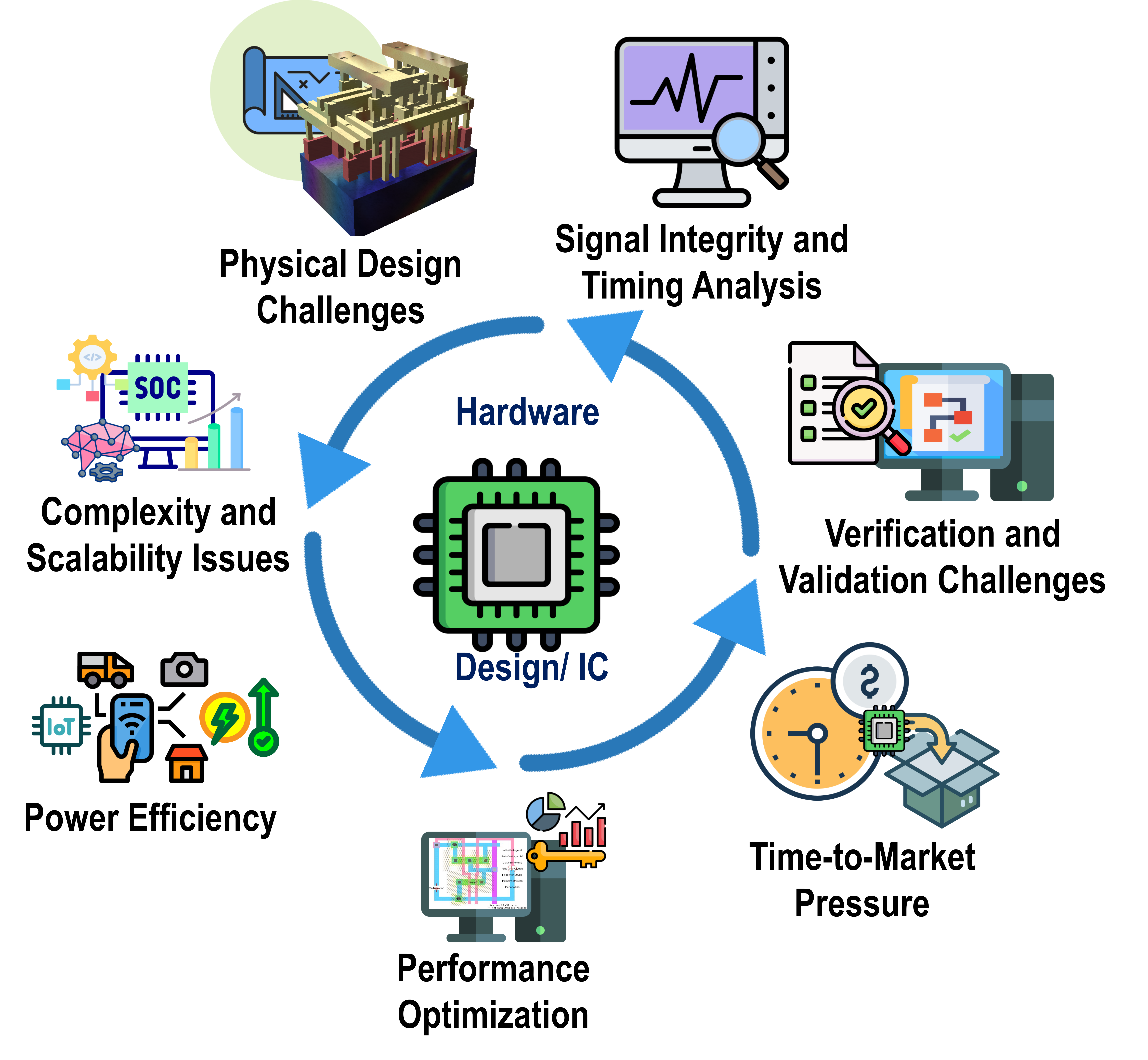}
    \caption{An overview of key challenges in hardware design: navigating Physical Design Challenges, ensuring strict Signal Integrity and Timing Analysis, overcoming  Verification and Validation requirements, managing Time-to-Market Pressures, executing Performance Optimization, maximizing Power Efficiency, and addressing the Complexity and Scalability Issues associated with IC architectures.} 
    \label{fig:challenges}
\end{figure}

The advancement in hardware design is progressing exponentially, leading to significant and unpredictable challenges that require continuous attention. Figure \ref{fig:challenges} illustrates the key challenges in the hardware design process, which include power efficiency, scalability, performance optimization, signal integrity, timing analysis, verification and validation, physical design constraints, and time-to-market pressures.
Modern Integrated Circuits (ICs) integrate billions of transistors with intricate interdependencies, making seamless functionality and scalability critical challenges. This increasing complexity has driven the development of System-on-Chip (SoC) architectures~\cite{tehranipoor2024large}, which address these challenges by compacting designs and enhancing performance. However, SoCs introduce new difficulties in verification and validation, as traditional methods struggle to handle the vast number of components and interactions within a single chip~\cite{forbes, yu2023machine}. Power management is a critical focus, especially with the proliferation of IoT and mobile devices. Designers must innovate to reduce power consumption and manage heat dissipation without compromising performance~\cite{10.1109/ASP-DAC58780.2024.10473904}. Performance optimization remains a primary goal in applications such as AI accelerators and high-speed communication systems, where achieving high-speed operations, low latency, and superior throughput are essential~\cite{liu2023verilogeval}. Signal integrity and timing analysis are equally vital. As signal paths shrink and frequencies rise in advanced nodes, ensuring reliable data transmission and synchronization has become increasingly challenging. Static Timing Analysis (STA) is commonly employed, but its accuracy is often hindered by the growing complexity and miniaturization of components~\cite{ye2023graph, chowdhury2022rapta, beheshti2023advanced}. Physical design challenges, such as floor planning, layout optimization, and routing, further complicate the hardware development process. Efficiently arranging chip components with proper routing to balance performance, power consumption, and thermal management is critical~\cite{yang2024floorplanning} and necessitates advanced algorithms and tools. Verification and validation have become increasingly resource-intensive and time-consuming, demanding advanced techniques to ensure all modules adhere to specifications and function as intended. Physical constraints in design also impact manufacturability and yield, requiring careful optimization of layouts and static timing. Time-to-market pressures add to these challenges, as delays in addressing any issue can lead to significant market disadvantages~\cite{fang2024assertllm}.

\begin{figure*}[!htp]
    \centering
    \includegraphics[width=0.80\linewidth]{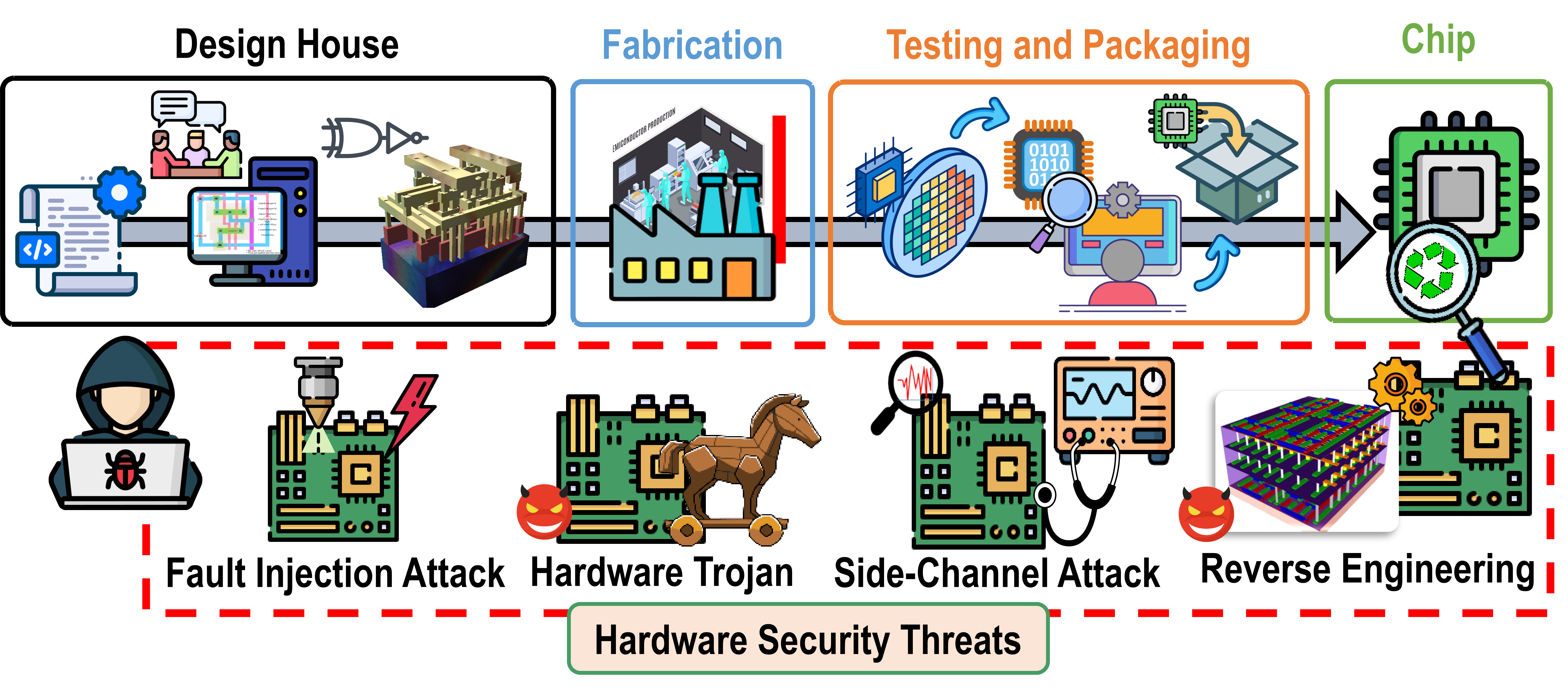}
    \caption{Overview of the entire IC supply chain, from the initial design house to the final chip deployment, and  highlighting the key Hardware Security threats possible during the design process, such as fault injection attacks, hardware Trojans, side-channel attacks, and reverse engineering.} 
    \label{security_threats}
\end{figure*}

ML can change traditional hardware design by increasing efficiency and accuracy at each stage of the workflow and by helping to solve challenges such as scalability, optimization, and decision-making that otherwise involve considerable manual effort and computational resources.
The selection and integration of third-party IPs can be optimized through ML models in the design specification and integration stage using compatibility and performance metrics prediction. Predictive ML techniques, for instance, analyze design requirements to provide suitable recommendations for IPs by reducing integration errors and improving design efficiency. Tools based on ML, such as ChipGPT~\cite{chang2023chipgptfarnaturallanguage}, already help designers at the RTL design stage with power-performance-area-optimized Verilog code. Similarly, ML models have been trained on hardware datasets to detect and repair logical errors in HDL designs, which helps enhance design reliability and scalability.
ML algorithms enhance optimization during logic synthesis and gate-level netlist generation by translating HDL into efficient gate-level netlists. In this respect, reinforcement learning methods optimize adaptive constraints for balanced timing, power, and area requirements~\cite{10361023}. ML significantly improves physical synthesis and layout, especially for placement and routing. Gao et al. demonstrated using graph-based neural networks to predict congestion hotspots and optimize wire length, reducing timing violations and ensuring manufacturability~\cite{10.1145/3489517.3530597}. Additionally, convolutional neural networks (CNNs) accelerate power grid performance prediction, offering faster evaluations than traditional methods.
Additionally, ML  can be used for testing and validation to automate test pattern generation and improve fault detection. Advanced ML models analyze test data to predict future failures, optimizing overall testing and fault coverage. These ML-driven improvements meet demands for complexity, scalability, and performance in modern hardware design with significantly reduced manual effort throughout the flow.

\subsection{Hardware Security and Challenges}

The evolution of hardware in recent years has introduced numerous security challenges that demand innovative solutions. Significant hardware challenges prevalent today include Trojans, IP theft, side-channel attacks, fault injection attacks, reverse engineering, and supply chain attacks~\cite{maynard2024reconfigurable, vishwakarma2024uncertainty, aghamohammadi2024lipstick, vishwakarma2023risk}. The globalization of the hardware supply chain has exacerbated these problems, necessitating robust detection and mitigation solutions~\cite{10443689}. Hardware Trojans are malicious and stealthy circuit alterations that compromise device integrity and functionality. The insertion of hardware Trojans is one of the most significant security challenges in hardware security, reported as one of the stealthiest forms of cyber attacks~\cite{ye2023graph}. Another critical attack is the side-channel attack, where attackers exploit circuit leakages, such as power consumption and EM emissions, to extract sensitive information from the system~\cite{10691788}. Fault injection attacks, where faults are deliberately induced in the circuit, represent another form of hardware attack, leading to data corruption and unauthorized access to the device's data and systems~\cite{10.1145/3649329.3657353}. Additionally, ICs are vulnerable to reverse engineering, which involves analyzing the IC to uncover its design and functions, preventing unauthorized replications, and identifying design security vulnerabilities~\cite{thakur2023autochip}.

Figure \ref{security_threats} outlines key hardware security threats across the IC design lifecycle, including hardware Trojans, side-channel attacks, fault injection, and reverse engineering.
Attackers try to exploit the IC for unintended leakages, such as power consumption or electromagnetic emissions, while the IC operates. 
In fault injection attacks, the attacker deliberately induces faults to corrupt data and disrupt systems. 
Conversely, reverse engineering attacks can apply to deployed chips. Such attacks allow the attacker to reconstruct designs and extract proprietary information from ICs during the testing and packaging stages. 

ML for hardware security has emerged as a research subject and effectively mitigates complex security challenges. One key area where ML techniques excel is hardware Trojan detection, with models trained to identify anomalies in circuit behavior that may indicate malicious modifications~\cite{sarihi2022hardware,cruz2022automatic}. Studies have shown that ML algorithms analyze power and other side-channel data to detect discrepancies caused by hardware Trojans~\cite{zantout2018hardware}. Additionally, ML-based techniques effectively mitigate side-channel vulnerabilities by understanding and reducing leakage patterns, enhancing the system's overall security posture~\cite{MUSHTAQ2020101524,10.1145/3386263.3407586,10.1007/978-3-030-94507-7_22, Liakos2023}. Logic obfuscation is another critical area in hardware security that uses ML to enhance circuit design security by introducing transformations that make reverse engineering significantly harder~\cite{ye2023graph,9770796}. Surveys further highlight ML's role in both physical and logic testing techniques for hardware Trojan detection~\cite{Mukherjee2022,10.1145/3579823}. ML also extends its applications to fault injection attack prevention and supply chain security, providing scalable, automated, and efficient solutions to reinforce hardware security~\cite{10.1145/3548606.3560690, LIAKOS2020103295, TAUHID2023100114}. By leveraging ML's capabilities, researchers and engineers are developing robust countermeasures to protect against a wide variety of hardware security threats.

\section{Attention-Based Hardware design}
\label{sec:AttentionBasedHardwareDesign}
In Section \ref{sec:background}, we covered the current challenges in hardware design. Now, we will delve into the LLM-based and other attention-based solutions identified for these challenges and examine ongoing research into new technologies to provide better solutions.

\begin{figure}[!t]
    \centering
    \includegraphics[width=1\linewidth]{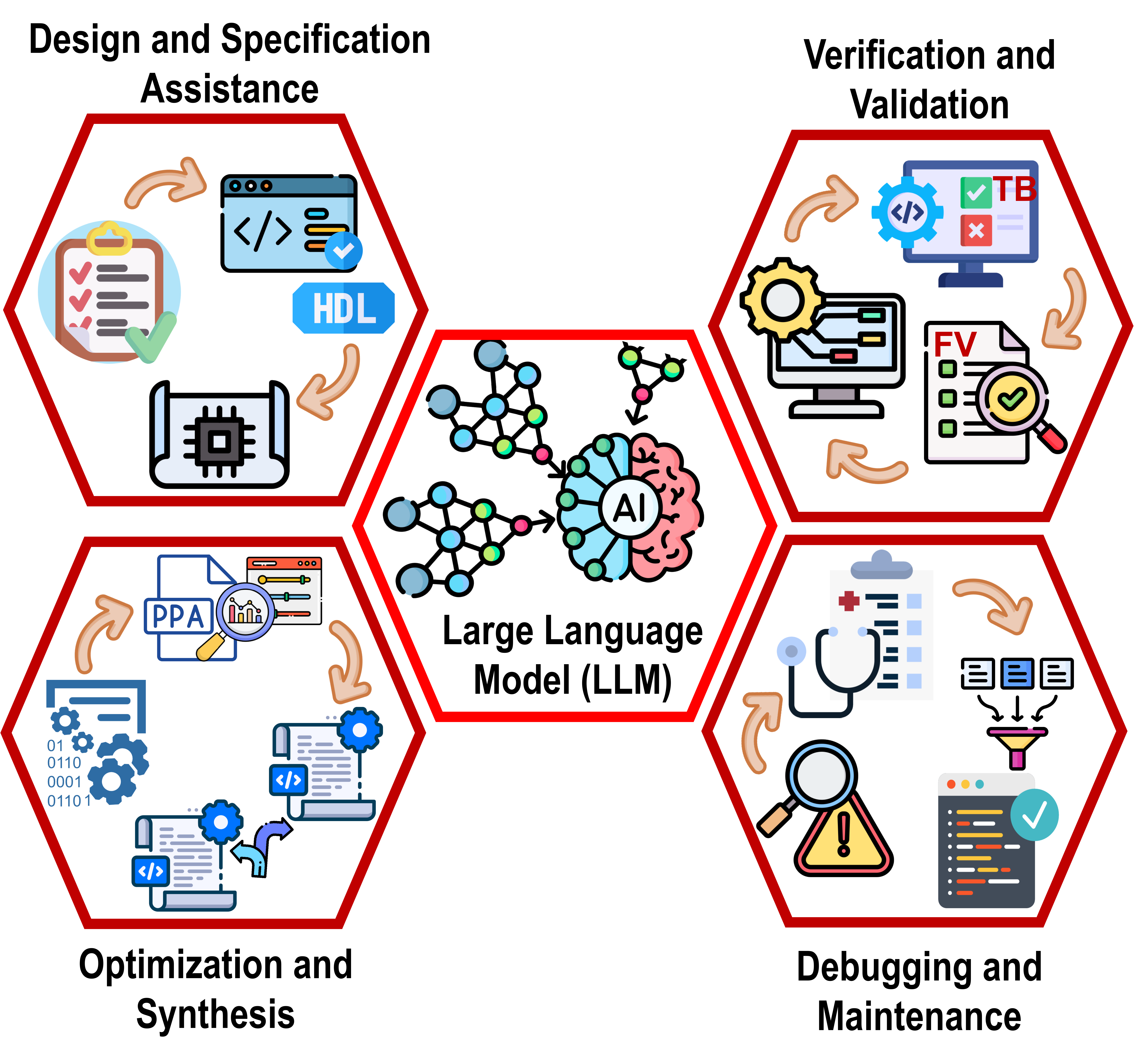}
    \caption{Overview of the critical applications of LLMs in digital hardware design, encompassing specification drafting, optimization, verification, and debugging to streamline the design lifecycle. Each application category includes key functionalities such as specification drafting, PPA optimization, testbench generation, and error correction.} 
    \label{fig:LLM}
\end{figure}

\subsection{LLM-based Digital Hardware Design}

This section explores the research that used LLMs for hardware design using various techniques.
Figure \ref{fig:LLM} illustrates the integrated overview of the critical applications of LLMs in digital hardware design. It indicates the processes in the hardware design flow that benefit most from LLMs. 
This paper categorizes the design process into four major aspects: design and specification assistance, optimization and synthesis, verification and validation, and debugging and maintenance. Each section indicates distinct hardware stages where LLMs ensure seamless integration, automation, and enhancement across these sections. LLMs are capable of supporting tasks such as specification drafting, Verilog code generation, performance optimization, testbench generation, and error detection. This figure provides a comprehensive overview, setting the stage for a deeper examination of each domain in the subsections that follow.

In this survey, we classify a method as LLM-based when its primary reasoning and generation interface relies on language or code token prediction, typically operating over natural-language prompts, HDL text, specifications, or documentation-to-code mappings. By contrast, we classify methods as broader attention-based hardware-design approaches when transformer or graph-attention mechanisms operate directly on circuit graphs, synthesis trajectories, placement states, or other structured design representations without primarily functioning as language models~\cite{li2024circuittransformerendtoendcircuit,yang2024floorplanning}. This distinction is important because several recent works customize transformer-based models for HDL generation while still preserving the language-modeling interface~\cite{liu2024rtlcoder,nadimi2024multiexpert}, whereas others use attention as a structured learning mechanism for placement, logic synthesis, or representation learning~\cite{li2024circuittransformerendtoendcircuit,yang2024floorplanning}. Within LLM-based digital hardware design, the literature broadly progresses along four recurring directions: specification-to-RTL generation, optimization and synthesis, verification and validation, and debugging and maintenance. Some works focus on translating natural-language intent into HDL, while others emphasize design refinement, correctness assurance, or post-generation repair. The choice of AI model is also closely tied to the target task: larger general-purpose models are commonly used for broad reasoning and multi-step planning, whereas smaller or fine-tuned hardware-specific models are preferred for efficient and syntax-sensitive HDL generation.

\subsubsection{Design And Specification Assistance}
\begin{figure}[!t]
    \centering
    \includegraphics[width=1\linewidth]{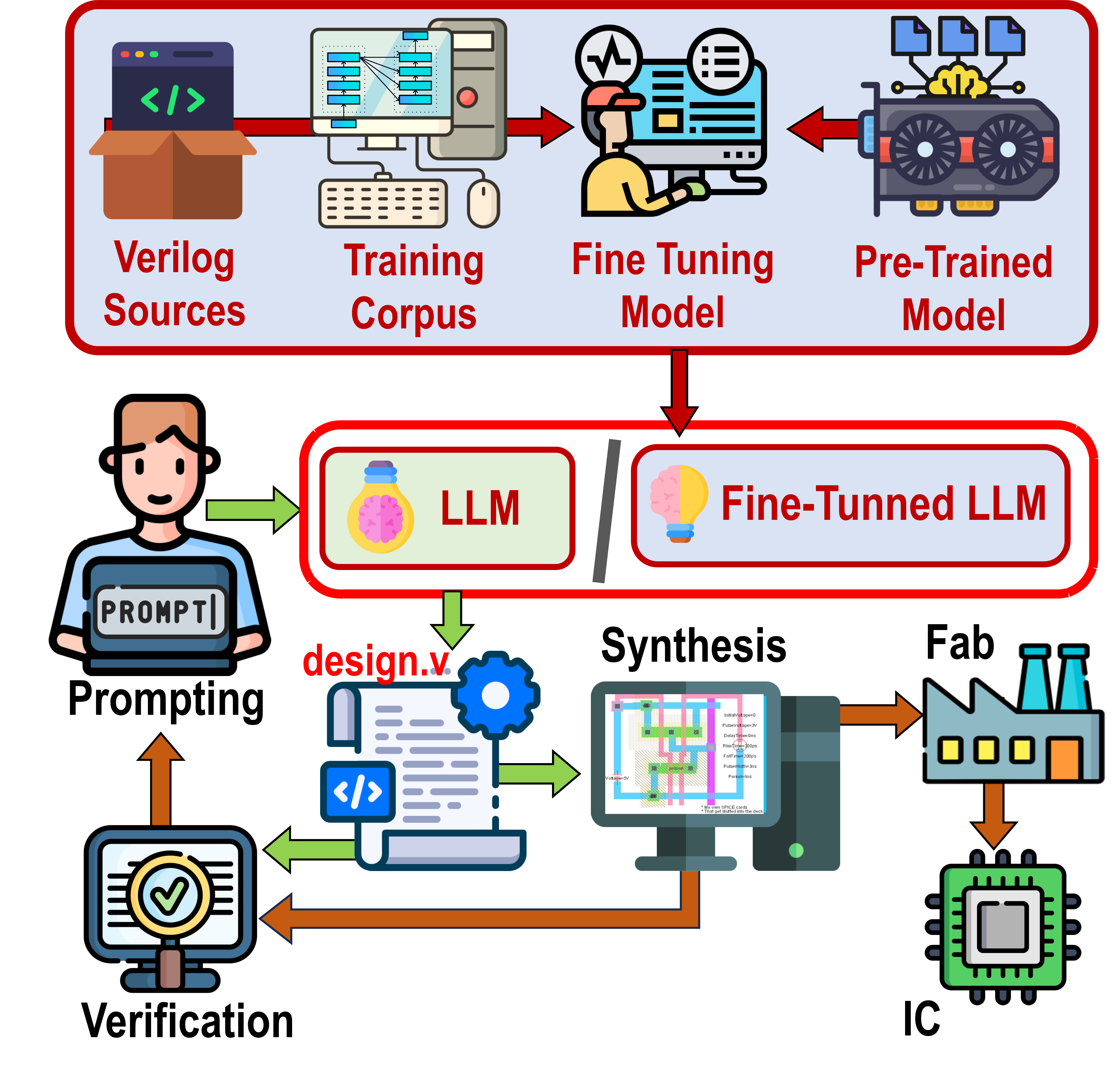}
    \caption{Workflow of hardware design using LLMs, highlighting how both prompting with pre-trained models and fine-tuning with Verilog sources and training data enable the complete hardware design process.} 
    \label{fig:llm1}
\end{figure}

In this category, LLMs aid in the initial stages of hardware design by drafting design specifications and converting natural language descriptions into formal HDLs. 
Figure \ref{fig:llm1} illustrates the comprehensive workflow of hardware design using LLMs, demonstrating their role in automating and enhancing various stages of the IC design process. The process begins with a human prompting in natural language, and then, depending on the type of LLMs, the output files as Verilog code are generated. This process enables efficient design and specification assistance in the chip design process. The generated Verilog codes are then synthesized and proceed to fabrication, with the verification steps in between to ensure correctness and compliance with performance requirements. The iterative feedback loop between the verification and synthesis stages highlights how LLMs optimize the overall design flow by reducing errors and improving efficiency at each stage of the design timeline.

To support this, a significant body of research has focused on leveraging LLMs for Verilog code generation. Thakur et al. pioneered this area by evaluating the ability of LLMs to generate syntactically correct and functionally accurate Verilog RTL code by fine-tuning models on a large Verilog corpus and benchmarking their performance with the fine-tuned CodeGen-16B model, achieving the highest accuracy~\cite{thakur2022benchmarking}. In another similar research, Thakur et al. compiled the largest corpus of Verilog code from different open-source repositories, which was then used to fine-tune an LLM. The resulting VeriGen model was evaluated against advanced general-purpose LLMs regarding the correctness of the generated HDL code, with the fine-tuned model demonstrating significantly improved performance~\cite{thakur2023verigen}. In another research endeavor, Thakur et al. further introduced Autochip, a fully automated feedback-driven Verilog code generation tool that iteratively refines designs based on compiler output~\cite{thakur2023autochip}. 

Further studies expanded the exploration of LLMs in functional HDL generation. Yang et al. highlighted the potential of GPT-4 for generating functional Verilog code for complex accelerator-style designs such as systolic arrays and AI accelerators for ResNet and MobileNet, although substantial manual optimization was still required for more complex designs~\cite{yang2023new}. Similarly, Yanık et al. introduced ShortCircuit, a tool that automates front-end digital IC design using ChatGPT and OpenLane. ShortCircuit significantly reduced design time for ASIC and FPGA implementations by generating HDL, GDS, and bitstream files~\cite{10382808}. Tomlinson et al. also leveraged ChatGPT to generate synthesizable and functional Verilog descriptions for a programmable Spiking Neuron Array ASIC through natural-language prompts, validating the design through simulation and preparing it for fabrication using SkyWater 130nm technology~\cite{tomlinson2024designing}.

Expanding beyond direct HDL generation, several studies focused on improving generation quality through specialized datasets, fine-tuning, retrieval augmentation, and task decomposition. Chang et al., in the “Data is all you need” paper, proposed an automated data augmentation framework for generating high-quality Verilog and EDA-script datasets by converting Verilog files into abstract syntax trees, applying predefined repair rules, and leveraging GPT-3.5 for EDA-script descriptions~\cite{chang2024data}. Fine-tuning Llama 2-7B and Llama 2-13B on these datasets achieved an 11.8\% improvement in pass rate over prior benchmarks~\cite{chang2024data}. 

Similarly, Liu et al. introduced RTLCoder, a lightweight and customized LLM for hardware design that outperformed GPT-3.5 while remaining efficient enough for local deployment through 4-bit quantization~\cite{liu2024rtlcoder}.In parallel, Nadimi and Zheng proposed the Multi-Expert Verilog LLM (MEV-LLM), which leveraged multiple fine-tuned LLMs specialized by design complexity to improve Verilog generation quality~\cite{nadimi2024multiexpert}. Evaluated on HDLBits tasks, the framework achieved up to a 23.9\% improvement in syntactically and functionally correct Verilog generation compared to prior models. Gao et al. further introduced AutoVCoder, which incorporated retrieval-augmented generation (RAG) to improve syntactic and functional correctness through knowledge retrieval and iterative refinement, significantly reducing hallucination-related errors~\cite{gao2024autovcodersystematicframeworkautomated}. 

To further refine HDL generation capabilities, Nakkab et al. explored hierarchical prompting techniques for handling complex hardware design tasks~\cite{nakkab2024romebuiltsinglestep}. Their framework decomposed hardware generation into structured subproblems, enabling efficient stepwise HDL generation. Experimental evaluations demonstrated that hierarchical prompting significantly improved generation quality for complex designs, including the successful generation of 16-bit MIPS and 32-bit RISC-V processors. 

Viewed together, these works reflect a clear progression rather than isolated improvements. Thakur et al. established strong baselines for Verilog generation by showing that hardware-specific fine-tuning substantially improves syntactic and functional correctness over general-purpose models~\cite{thakur2022benchmarking,thakur2023verigen}. Yang et al. extended this direction to more complex accelerator-style designs, exposing a key limitation of early RTL-generation systems: promising results on smaller designs do not necessarily generalize to more complex architectures without substantial manual refinement~\cite{yang2023new}. Subsequent works then addressed complementary bottlenecks through curated datasets, retrieval augmentation, hierarchical prompting, and specialized model architectures~\cite{chang2024data,gao2024autovcodersystematicframeworkautomated,nadimi2024multiexpert,nakkab2024romebuiltsinglestep}. In this sense, the literature evolves along several complementary axes, including domain adaptation, task complexity, dataset quality, and structured refinement.


A defining trend in 2025 is the shift from purely supervised fine-tuning toward reasoning-augmented and reinforcement-learning (RL)-based RTL generation. Wang et al. proposed VeriReason, which couples supervised fine-tuning with Group Relative Policy Optimization (GRPO) and testbench feedback so that the model learns to reason about and self-check its Verilog, reaching 83.1\% functional correctness on VerilogEval and up to a 2.8$\times$ increase in first-attempt correctness over strong baselines~\cite{wang2025verireason}. Teng et al. extended this direction with VeriRL, curating the large Veribench-53K corpus and introducing a sample-balanced weighting strategy to mitigate the catastrophic forgetting and reward sparsity that often destabilize RL fine-tuning for HDL~\cite{teng2025verirl}. Complementing reasoning with structural awareness, Zhao et al. augmented SFT with RL while embedding Abstract Syntax Trees (ASTs) and Data Flow Graphs (DFGs) into the reward signal to preserve syntactic integrity and semantic dependencies, yielding a 12.57\% improvement on VerilogEval-Human~\cite{zhao2025enhancing}. In parallel, data-centric efforts matured: Wei et al. introduced VeriCoder, fine-tuned on more than 125{,}000 functionally validated RTL examples produced through a generate--simulate--repair loop, improving functional correctness by up to 71.7\% on VerilogEval and 27.4\% on RTLLM~\cite{wei2025vericoder}. Together, these works mark a transition from correctness-by-imitation toward correctness-by-feedback, in which explicit reasoning, structural priors, and execution signals are folded directly into training.

Beyond direct HDL generation, another emerging direction positions LLMs as an interaction layer between designers and the broader hardware design environment. Rather than simply generating Verilog from prompts, these systems interpret natural-language queries, retrieve structured design information, translate user intent into tool-compatible actions, and return results in a form that is easier for designers to use.
Abdelatty and Reda introduced HDLCopilot, an LLM-powered query system enabling natural-language interaction with Process Design Kits (PDKs)~\cite{abdelatty2024hdlcopilothardwaredesignlibrary}. The framework translates user queries into SQL statements for retrieving information from a PDK database and then produces natural-language responses. Its four-agent architecture---Dispatcher, Selector, SQL Generator, and Interpreter---supports different stages of query understanding and execution. Evaluated on the SkyWater 130 nm PDK database, HDLCopilot achieved 94.23\% accuracy and 98.07\% efficiency in information retrieval. 

A related direction extends LLM-based design assistance beyond text-only interaction by incorporating multimodal inputs. Chang et al. proposed a multimodal generative AI framework combining natural-language descriptions with visual input to improve Verilog generation, particularly for designs involving complex spatial relationships and module interactions~\cite{chang2024naturallanguageenoughbenchmarking}. Their study demonstrated multimodal input substantially improved both syntactic correctness and functional performance, with testbench passing rates increasing from 46.88\% to 71.81\% for GPT-4-based models and from 13.41\% to 25.88\% for LLaMA-based models. These findings suggest that multimodal context can reduce ambiguity and better capture design intent for tasks involving diagrams, state machines, and multi-module hardware structures.~\cite{chang2024naturallanguageenoughbenchmarking}. 

Recent works have also explored broader specification and design-assistance applications. Mengming et al. demonstrated how LLMs can streamline IC development by assisting with architecture specification generation, from high-level intent to detailed RTL-oriented descriptions~\cite{li2024specllm}. Zhao et al. introduced CodeV, a framework enhancing Verilog generation through multi-level summarization and high-quality instruction tuning~\cite{zhao2024codevempoweringllmsverilog}. Fine-tuned CodeLlama variants achieved strong improvements over GPT-4 and BetterV on benchmark tasks. Complementarily, DeLorenzo et al. proposed CreativEval, a framework for evaluating creativity in LLM-generated hardware designs across dimensions such as fluency, flexibility, originality, and elaboration~\cite{delorenzo2024creativeval}. Their study showed GPT-3.5 emerging as one of the most creative models among evaluated systems.

As these systems evolved, several studies expanded toward workflow-level automation and broader EDA integration. Wang et al. introduced ChatPattern, an LLM-driven framework for generating and customizing layout patterns using natural-language instructions~\cite{10.1145/3649329.3657361}. The framework autonomously decomposes tasks and operates design tools to generate layout patterns while supporting pattern modification and conditional generation. Liu et al. similarly leveraged Llama2-70B to automate the layout of silicon photonic devices, demonstrating over 99\% success in generating correct layout code from natural-language prompts~\cite{10.1007/978-3-031-63378-2_72}. 

He et al. further proposed ChatEDA, an autonomous EDA agent powered by the AutoMage large language model~\cite{he2024chatedalargelanguagemodel}. ChatEDA automates the RTL-to-GDSII workflow through task planning, script generation, and execution management, outperforming GPT-4 in several task-planning benchmarks. 
These workflow-oriented systems differ in both scope and level of automation. HDLCopilot focuses primarily on natural-language interaction with structured design resources such as PDKs~\cite{abdelatty2024hdlcopilothardwaredesignlibrary}, whereas SpecLLM emphasizes architecture-specification assistance and textual summarization~\cite{li2024specllm}. By contrast, ChatEDA pushes toward fuller workflow orchestration by coordinating planning, script generation, and execution across multiple EDA stages~\cite{he2024chatedalargelanguagemodel}. Collectively, these studies illustrate the gradual evolution of LLMs from isolated HDL-generation tools toward broader design-assistance and workflow-orchestration frameworks.

The use of LLMs in high-level synthesis (HLS) addresses a different bottleneck from direct RTL generation. Rather than generating HDL directly, HLS-oriented approaches focus on bridging the gap between software-style C/C++ descriptions and synthesizable hardware implementations, reducing the manual effort required to adapt conventional software code for hardware synthesis.
HLS tools facilitate rapid hardware design from C code but face limitations due to incompatible code constructs. Collini et al. investigated the use of LLMs to refactor standard C code into HLS-compatible formats~\cite{collini2024c2hlsc}. Their study demonstrated iterative transformation of software implementations such as AES-128, QuickSort, and NIST randomness tests into synthesizable HLS-compatible code using compiler-guided refinement. Similarly, Meech introduced a methodology combining HLS tools and LLMs to generate, simulate, and deploy hardware designs using Amaranth HDL scripts generated through Microsoft Bing Chat~\cite{meech2024leveraginghighlevelsynthesislarge}. The framework successfully deployed a random number generator accelerator on an Icebreaker FPGA integrated with the Caravel SoC platform, highlighting the potential of LLM-assisted HLS for rapid prototyping of domain-specific accelerators.


The HLS direction advanced rapidly in 2025 with the arrival of dedicated datasets and specialized models. Khan et al. proposed SAGE-HLS, the first LLM fine-tuned specifically for HLS code generation, applying an AST-guided procedure over a 16.7K-sample dataset built by porting Verilog to C/C++, and reporting near-100\% synthesizability with 75\% functional correctness~\cite{khan2025sagehls}. Zou et al. released HLStrans, a benchmark-scale dataset of more than 124{,}000 paired C and HLS programs with full testbenches and synthesis-based annotations, showing that retrieval and fine-tuning substantially improve C-to-HLS success rates~\cite{zou2025hlstrans}. These resources begin to address the long-standing data scarcity that has limited LLM-assisted HLS.


Recent work has also begun to question the implicit assumption that an HDL is always the right generation target. Tasnia and Rahman proposed OPL4GPT, an application-space exploration that tasks GPT-4o with implementing the same System-on-Chip functional blocks in both C++ (via high-level synthesis) and Verilog, using a closed-loop syntax- and functionality-feedback mechanism~\cite{tasnia2025opl4gpt}. Their study shows that the optimal input language is task-dependent: data-intensive blocks such as cryptographic or AI accelerators reach higher functional correctness when generated in C++, whereas fine-grained structural components are better expressed directly in Verilog. This reframes language selection itself as a design decision that LLM-assisted flows should expose rather than fix in advance.

Collectively, these studies indicate that improvements in LLM-based hardware design do not arise from a single innovation, but from a recurring set of mechanisms. Hardware-specific fine-tuning improves familiarity with HDL syntax and design structure; curated datasets substantially improve syntactic correctness and functional pass rates; hierarchical prompting and task decomposition make complex hardware generation more tractable; and retrieval, compiler feedback, and iterative refinement reduce hallucinations by incorporating external correctness signals~\cite{chang2024data,gao2024autovcodersystematicframeworkautomated,nakkab2024romebuiltsinglestep}. More broadly, the literature demonstrates a progression from isolated prompt-based HDL generation toward multimodal design assistance, workflow orchestration, and increasingly integrated hardware-design automation environments.

\subsubsection{Optimization And Synthesis}
Optimization and synthesis have emerged as important areas in which LLMs can improve both the efficiency and quality of hardware design workflows. In this context, these models are used to assist in generating HDL from high-level specifications and to refine design quality, support synthesis decisions, and improve downstream implementation outcomes. Importantly, reported improvements in power, performance, and area (PPA) arise through several distinct pathways that should not be conflated. Some methods make generation itself more constraint-aware by conditioning the model on design objectives or target quality metrics. Others operate through post-generation refinement, search, or repair, in which an initial RTL candidate is improved iteratively based on external feedback. A third class acts downstream of code generation by optimizing synthesis recipes, backend parameters, or design-space exploration loops using prior design knowledge and tool outputs. Distinguishing these pathways is important because not all reported PPA gains come from better HDL generation alone; in many cases, the improvement is achieved through search, retrieval, reinforcement learning, or tighter integration with EDA workflows after the model’s initial output has been produced~\cite{anonymous2023improving,chowdhury2024retrievalguided,jiang2024iicpilotintelligentintegratedcircuit,thorat2023advanced}.

A common thread across several studies is the use of LLMs to improve PPA metrics through specialized frameworks and optimization methodologies. For instance, Thorat et al. introduced the VeriPPA framework to optimize PPA constraints in Verilog designs, significantly improving the quality of LLM-generated Verilog code and bringing it closer to industry expectations~\cite{thorat2023advanced}. Similarly, Kaiyan Chang et al. introduced ChipGPT, a framework that improves hardware quality through a post-LLM search strategy. By integrating design-space exploration into the workflow, ChipGPT improves PPA metrics beyond what can be achieved through direct generation alone~\cite{anonymous2023improving}. 

Chowdhury et al. further advanced this direction through the ABC-RL methodology, which leverages prior design data and a retrieval-guided reinforcement learning approach to optimize logic synthesis, achieving up to 24.8\% improvement in Quality-of-Result (QoR) and up to 9$\times$ runtime reduction compared to state-of-the-art techniques~\cite{chowdhury2024retrievalguided}. In this framework, attention is used to encode synthesis recipes in a form that enables the remaining learning components to better model and predict the circuit state resulting from each synthesis optimization step. 

Another important direction couples HDL/RTL generation with search-based refinement and synthesis-aware optimization to improve both functional correctness and implementation quality. DeLorenzo et al. proposed an approach that integrates LLMs with Monte Carlo Tree Search (MCTS) to generate high-quality RTL code~\cite{delorenzo2024make}. Rather than committing to a single decoding path, MCTS explores multiple candidate continuations and evaluates partial designs using external signals such as compilation success, simulation outcomes, logical constraints, or synthesis-oriented quality metrics. This structured exploration helps the model recover from locally plausible but globally poor choices and increases the likelihood of producing functionally correct and higher-quality hardware artifacts~\cite{delorenzo2024make,li2024circuittransformerendtoendcircuit}. Their transformer-decoding algorithm, guided by MCTS, produces compilable and functionally correct RTL code while significantly improving the area-delay product (ADP) for a range of hardware modules.

Nazzal et al. introduced the Systolic Array-based Accelerator DataSet (SA-DS), designed to facilitate the application of LLMs to the generation and optimization of DNN hardware accelerators~\cite{nazzal2024dataset}. Created using the Gemmini generator, the dataset includes diverse spatial-array configurations paired with verbal descriptions and Chisel code. The study shows that fine-tuning LLMs on SA-DS improves both the quality and efficiency of hardware design generation, highlighting the potential for reduced human intervention in accelerator design workflows.

As design complexity increases, recent research is also moving beyond single-model prompting toward autonomous multi-agent frameworks that coordinate specialized reasoning modules across the design flow. The ChipMind architecture exemplifies this shift by deploying specialized LLM agents for digital, analog, and security workflows under the orchestration of a central Coordinator Agent~\cite{firouzi2025chipmnd}. In the digital flow, dedicated agents handle HDL generation, PPA-aware synthesis, and functional verification. The Coordinator further leverages a Knowledge Graph and Knowledge Base to retrieve optimization strategies and iteratively refine artifacts until physical and timing constraints are satisfied.


This multi-agent paradigm broadened considerably in 2025. Wu et al. proposed EDAid, a collaboration system in which agents driven by fine-tuned ChipLlama experts converge on EDA-flow automation and outperform single-agent baselines on complex tool-calling chains~\cite{wu2025edaid}. Lu et al. introduced AutoEDA, a microservice- and MCP-based agent framework that controls RTL-to-GDSII flows from natural language, reporting up to 9.9$\times$ higher task accuracy while cutting token usage by roughly 97\% relative to in-context learning~\cite{lu2025autoeda}. Jiang et al. extended agentic coordination to hardware/software co-design with MACO, an open-source multi-agent framework for coarse-grained reconfigurable arrays (CGRAs) that pairs architecture exploration with self-learning error correction~\cite{jiang2025maco}. These systems indicate a clear movement from single-model assistants toward orchestrated agent ecosystems spanning the full design flow.

Beyond RTL-level optimization, researchers are also integrating LLMs with EDA tools to improve backend design scalability and usability. Jiang et al. introduced IICPilot, an intelligent IC backend design system that uses LLMs and a multi-agent framework to automate various backend design processes using open-source EDA tools~\cite{jiang2024iicpilotintelligentintegratedcircuit}. IICPilot automates script generation, EDA tool invocation, design space exploration of parameters, and dynamic resource allocation through containerization. The framework separates the backend workflow from specific EDA tools through a unified interface and leverages a LangChain-based multi-agent system for tasks such as floorplanning and routing. Experimental validation on platforms like iEDA and OpenROAD demonstrated significant improvements in task completion time and resource utilization, achieving up to 32.75\% optimization in PPA metrics. The system's scalability and fault tolerance were further evaluated on a multi-node setup using 400 OpenCores benchmarks.

Ho and Ren further extended the application of LLMs to physical layout optimization, specifically targeting advanced semiconductor nodes at the 2nm technology node~\cite{ho2024largelanguagemodelllm}. Their methodology employs the ReAct prompting technique to integrate human design expertise with LLM reasoning, resulting in up to 19.4\% smaller cell area and 23.5\% more LVS/DRC clean layouts compared to traditional methods. This work underscores the versatility of LLMs in understanding and optimizing netlist topologies, device clustering constraints, and physical layouts.

Two 2025 directions further illustrate how optimization is moving away from one-shot generation. Abdollahi et al. introduced RocketPPA, a code-level power, performance, and area (PPA) predictor that operates directly on HDL using an LLM-based regressor coupled with a mixture-of-experts (MoE) architecture, improving area and power prediction accuracy by 13.6\% and 14.7\%, respectively, while delivering more than a 20$\times$ speedup over prior estimators~\cite{abdollahi2025rocketppa}; fast code-level PPA feedback of this kind is exactly what closes the loop for search- and RL-based generation. Krupp et al. approached optimization conversationally with CRADLE, a multi-agent design-space-exploration framework in which generator and critic agents iteratively refine RTL toward FPGA-resource targets, reducing LUT and flip-flop usage by 48\% and 40\% on RTLLM benchmarks through self-verification and correction~\cite{krupp2025cradle}. These systems reinforce the survey's broader observation that the strongest optimization results arise when LLM reasoning is paired with external, quantitative feedback rather than from generation quality alone.

Overall, optimization-oriented LLM methods differ from basic HDL-generation systems because their objective is not only to produce syntactically valid code but also to improve implementation quality under design constraints. Existing work shows three main routes toward this goal: constraint-aware generation, post-generation search and refinement, and downstream EDA-flow optimization. As a result, the strongest optimization frameworks increasingly combine LLM reasoning with external feedback from synthesis tools, reinforcement learning, retrieval mechanisms, or multi-agent workflow orchestration.

\subsubsection{Verification And Validation}
Verification and validation are crucial in hardware design to ensure functionality and compliance with specifications. LLMs can assist by generating test cases, checking HDL correctness, producing hardware assertions, and supporting formal verification, thereby helping identify design flaws earlier in the development process. Across the verification literature, comparisons are most meaningful when viewed through a common set of evaluation dimensions. These include syntactic correctness, functional correctness, assertion or testbench quality, coverage, repair success, and the extent to which results are validated through simulation, formal methods, or silicon implementation. Not all papers report the same metrics, and some emphasize benchmark pass rates while others focus on coverage or proof-oriented validation. Making these dimensions explicit is important because improvements in one metric, such as syntax correctness, do not necessarily imply improvements in deeper functional or formal reliability.

A key area of focus has been the development of benchmarking frameworks to systematically evaluate the capabilities of LLMs in hardware design and verification. Lu et al. introduced  RTLLM as an open-source benchmark for systematic and quantitative evaluation of RTL design generation from natural language~\cite{10.1109/ASP-DAC58780.2024.10473904}. RTLLM focuses on syntax correctness, functional accuracy, and design quality, and also explores a self-planning prompting technique to improve LLM performance. Similarly, Mingjie et al. developed VerilogEval, a benchmarking framework for automatic functional correctness testing of generated Verilog code~\cite{liu2023verilogeval}.  Their dataset was derived from HDLBits, an online coding platform for hardware engineers, and their study showed that supervised fine-tuning (SFT) with synthetic problem-code pairs can enhance the Verilog generation capabilities of pre-trained LLMs.

Another prominent theme is the use of LLMs for assertion generation and formal verification, which are essential for ensuring design reliability. Fang et al. proposed AssertLLM, demonstrating how LLMs can generate and evaluate hardware verification assertions to improve design reliability~\cite{fang2024assertllm}. Complementing this direction, Mali et al. introduced ChiRAAG, a framework that automates the generation of SystemVerilog Assertions (SVAs) from natural-language specifications using LLM prompting~\cite{mali2024chiraag}. This work helps streamline the formal property verification process in hardware design. Further advancing this area, Hassan et al. proposed a formal verification methodology that uses GPT-4 to automate invariant generation and integrates mutation testing to validate these invariants for hardware designs~\cite{10546729}. The method was demonstrated on the ISCAS-85 C432 27-channel interrupt controller, achieving effective invariant validation and a mutation score of 1 out of 20 mutants.
 
The automation of testbench generation and functional verification has also been a significant focus. Qiu et al. introduced AutoBench, a systematic and generic framework for generating Verilog testbenches using LLMs for RTL verification~\cite{qiu2024autobenchautomatictestbenchgeneration}. AutoBench combined Python and Verilog codes in a hybrid testbench architecture, enhancing efficiency and effectiveness. The framework included comprehensive code generation, scenario checking, standardization, and automatic debugging. Its AutoEval component evaluates generated test benches using multiple metrics, demonstrating higher pass ratios and task completion rates than traditional methods. 
Similarly, Blocklove et al. evaluated the ability of LLMs such as ChatGPT-4 and ChatGPT-3.5 to generate functional Verilog code and corresponding test benches for hardware design and verification~\cite{10691811}. The study developed benchmarks to assess LLMs' performance across various hardware functions and validated the designs through simulation and silicon testing. Results showed that while LLMs can generate functional HDL code, their ability to produce comprehensive testbenches is limited, with ChatGPT-4 outperforming ChatGPT-3.5. The authors emphasize the need for improvements in error understanding, testbench generation, and handling complex designs. The author plans to further enhance dataset quality, training schemes, and integration with EDA tools to advance the automation of the digital design pipeline.

Beyond general-purpose frameworks, researchers have also investigated the use of LLMs for application-specific hardware design and validation. Vitolo et al. explored the use of ChatGPT-4 to automate the generation and validation of Verilog code for a Recurrent Spiking Neural Network (RSNN)~\cite{vitolo2024natural}. The study demonstrated the successful design of a 3x3x3 RSNN and corresponding test benches through an iterative LLM-assisted process. The resulting design was validated on tasks such as exclusive OR, IRIS flower classification, and MNIST handwritten digit classification and was further prototyped on an FPGA and implemented using SkyWater 130 nm technology. This work underscores the feasibility of using LLMs for complex hardware design, suggesting future research to enhance dataset quality, training schemes, and integration with electronic design automation tools for more sophisticated hardware projects.

LLM-assisted verification has also been connected to debugging, repair, and EDA integration. Thorat et al. introduced the veriRectify process, which uses simulator error diagnostics to improve the integrity correctness of generated Verilog code~\cite{thorat2023advanced}. In a related direction, Lily Jiaxin Wan et al. explored techniques such as quantization, pruning, and software/hardware co-design to optimize LLMs for efficient inference in circuit-design workflows~\cite{10473893}. They applied these optimized models to functional verification and introduced the Chrysalis dataset to support LLM-based debugging tools, aligning with broader efforts to integrate LLM capabilities into existing EDA environments.

Overall, verification-oriented LLM research has progressed from basic correctness evaluation toward more specialized support for assertions, testbenches, formal reasoning, and debugging. Benchmarking efforts such as RTLLM and VerilogEval provide a foundation for measuring syntactic and functional correctness, while assertion-generation and testbench-generation frameworks extend LLM usage into practical verification workflows. However, the literature also shows that verification quality cannot be judged by code generation success alone. Stronger validation requires coverage analysis, simulation evidence, formal checks, repair success, and, where possible, silicon or FPGA-level testing.

\subsubsection{Debugging And Maintenance}
LLMs may play a crucial role in debugging and maintaining hardware designs. They can assist in identifying and fixing syntax errors in HDL code, interpreting simulator and synthesis diagnostics, streamlining the debugging process, and improving the correctness of hardware designs. These capabilities facilitate more efficient troubleshooting and maintenance, especially when LLMs are integrated with compiler feedback, retrieved documentation, or EDA tool outputs.

Researchers have explored several directions for LLM-assisted debugging and maintenance. For instance, YunDa et al. developed RTLFixer, a framework that automates syntax correction in Verilog code using LLMs~\cite{10.1145/3649329.3657353}. RTLFixer uses Retrieval-Augmented Generation (RAG) and ReAct prompting, enabling LLMs to act as autonomous agents that interactively debug code using external feedback. Similarly, Yao et al. introduced HDLdebugger, which uses RAG to retrieve relevant HDL documentation and code examples to improve the debugging capability of LLMs~\cite{yao2024hdldebugger}. These works show that retrieval and tool feedback can make LLM-based debugging more reliable than direct prompt-based repair alone.

Another shared focus is improving the interpretability of error messages and making hardware design tools more accessible to novice designers. Qiu et al. explored this direction by using LLMs to generate user-friendly explanations for compile-time synthesis error messages from EDA tools such as Quartus Prime and Vivado~\cite{qiu2024explaining}. Their results showed that approximately 71\% of LLM-generated explanations were correct and complete, suggesting that LLMs can help bridge the gap between complex hardware tool outputs and designer understanding, especially in educational settings.

In addition to general debugging, LLMs have been applied to improve the reliability of HDL code generation for specific hardware applications. Xiang et al. investigated this direction by developing a three-phase Pulse Width Modulation (PWM) generator using LLMs~\cite{xiang2024digital}. Their study emphasized role specification, hierarchical design, and error-feedback mechanisms to address both syntax errors and high-level semantic challenges in LLM-generated code. The successful fabrication and validation of the PWM generator using tools such as iVerilog and GTKWave underscore the potential of LLMs in integrated circuit (IC) design. This work also highlights the importance of integrating LLMs with existing EDA tools to improve their practical utility.

Furthermore, LLMs have shown potential in repairing security-relevant hardware bugs. Ahmad et al. investigated the use of LLMs for automatically repairing security-relevant bugs in Verilog code~\cite{10462177}. Their study introduced a quantitative framework for evaluating LLM-based bug repair and compared LLM performance against state-of-the-art automated repair tools. Experimental results showed that LLMs, particularly GPT-4, can repair simple hardware security bugs and sometimes outperform existing tools. Detailed prompt engineering also improved repair success rates. However, LLMs struggled with more complex issues such as race conditions and multi-line fixes, suggesting that future work should explore hybrid methods, hardware-specific fine-tuning, stronger prompt strategies, and simulation- or formal-verification-guided repair.


The 2025 literature substantially expanded LLM-based hardware debugging along three axes: benchmarking, localization, and retrieval-augmented repair. On benchmarking, Li et al. introduced FixBench-RTL, a suite of buggy/corrected RTL pairs with natural-language bug descriptions and simulation testbenches spanning syntactic, functional, and security-relevant defects, showing that even popular LLMs still fall well short of practical debugging requirements~\cite{li2025fixbench}. On localization, Ahmad et al. proposed FLAG, which regenerates each line of RTL and compares it against the original to flag anomalies without any synthesis or simulation, identifying 38 of 120 real-world bugs within its top-five candidate locations~\cite{ahmad2025flag}. Wang et al. unified detection, localization, and repair in VeriDebug through contrastive embedding and guided correction, raising bug-fix accuracy to 64.7\% versus 36.6\% for GPT-3.5-turbo and 11.3\% for open-source baselines~\cite{wang2025veridebug}. Retrieval-augmented repair also matured: Qayyum et al. combined retrieval-augmented generation with LLM-based patching for Verilog HDL to improve identification and correction accuracy over prompt-only repair~\cite{qayyum2025llm}, and complementary studies continue to probe how prompting strategy and iterative feedback shape RTL bug-repair success~\cite{elnaggar2025adding}. Collectively, these works confirm the survey's thesis that LLMs become reliable debugging aids only when embedded in feedback- and retrieval-driven loops rather than used as standalone repair tools.

\begin{table*}[!t]
\centering
\caption{LLM-Based Digital Hardware Design}
\label{LLM_design}
\scriptsize
\renewcommand{\arraystretch}{1.15}
\setlength{\tabcolsep}{3pt}

\begin{tabularx}{\textwidth}{|
>{\centering\arraybackslash}m{0.7cm}|
>{\raggedright\arraybackslash}p{2.4cm}|
>{\centering\arraybackslash}m{1.15cm}|
>{\centering\arraybackslash}m{1.25cm}|
>{\centering\arraybackslash}m{1.25cm}|
>{\centering\arraybackslash}m{1.35cm}|
>{\raggedright\arraybackslash}X|}
\hline
\rowcolor[HTML]{E3E3E3}
\textbf{Year} &
\textbf{Method Name} &
\textbf{Design Assistance} &
\textbf{Optimization Synthesis} &
\textbf{Verification Validation} &
\textbf{Debugging Maintenance} &
\textbf{AI Model} \\
\hline

2022 &~\cite{thakur2022benchmarking} & \checkmark & & \checkmark & & CodeGen(2B, 6B, 16B), MegatronLM-355M, Code-davinci-002\\ \hline
2023 & ChipChat~\cite{10299874} & \checkmark & & & & ChatGPT-4\\ \hline
2023 & VerilogEval~\cite{liu2023verilogeval} & \checkmark & & \checkmark & & GPT(3.5, 4), CodeGen-16B-verilog\\ \hline
2023 & RTLLM~\cite{10.1109/ASP-DAC58780.2024.10473904} & & & \checkmark & & GPT-3.5\\ \hline
2023 &~\cite{thorat2023advanced} & \checkmark & \checkmark & \checkmark & & GPT(3.5, 4)\\ \hline
2023 & RTLFixer~\cite{10.1145/3649329.3657353} & & & & \checkmark & GPT(3.5, 4)\\ \hline
2023 & RTLCoder 2023~\cite{10691788} & & & \checkmark & & Mistral-7B, DeepSeek-Coder-6.7B\\ \hline
2023 & AutoChip~\cite{thakur2023autochip} & \checkmark & & & & GPT(3.5-turbo, 4), Claude 2, Code Llama 2\\ \hline
2023 & ShortCircuit~\cite{10382808} & \checkmark & & & & ChatGPT\\ \hline
2023 & VeriGen~\cite{thakur2023verigen} & \checkmark & & & & MegatronLM-355M, CodeGen(2B, 6B, 16B), J1-Large-7B, code-davinci-002, GPT(3.5-turbo, 4), PALM2\\ \hline
2023 & LLM4EDA~\cite{zhong2023llm4eda} & & \checkmark & & & GPT(3.5, 4), LLaMA2\\ \hline

2024 & LLM-AID~\cite{firouzi2024llm} & \checkmark & \checkmark & & & GPT-4\\ \hline
2024 & SpecLLM~\cite{li2024specllm} & \checkmark & & & & GPT-4\\ \hline
2024 & AssertLLM~\cite{fang2024assertllm} & & & \checkmark & & GPT(3.5, 4, 4-Turbo)\\ \hline
2024 & ChIRAAG~\cite{mali2024chiraag} & & & \checkmark & & GPT-4 Turbo\\ \hline
2024 &~\cite{delorenzo2024make} & & & \checkmark & & VeriGen-2B, fine-tuned version of CodeGen LLM\\ \hline
2024 &~\cite{10546729} & & \checkmark & & & GPT-4\\ \hline
2024 &~\cite{tomlinson2024designing} & \checkmark & & & & ChatGPT-4\\ \hline
2024 &~\cite{10473893} & & & \checkmark & & GPT-4, LLaMA2\\ \hline
2024 & HDLdebugger~\cite{yao2024hdldebugger} & & & & \checkmark & GPT-4, RTLFixer, VeriGen, CodeLlama-13B\\ \hline
2024 &~\cite{chang2024data} & & & & \checkmark & Llama 2-13B, Llama 2-7B\\ \hline
2024 & ChatPattern~\cite{10.1145/3649329.3657361} & \checkmark & & & & Not specified\\ \hline
2024 &~\cite{anonymous2023improving} & & & & \checkmark & GPT(3.5-Turbo, 4)\\ \hline
2024 &~\cite{delorenzo2024creativeval} & \checkmark & & & & GPT(3.5, 4), CodeLlama(7B, 13B), VeriGen(6B, 16B)\\ \hline
2024 &~\cite{nazzal2024dataset} & & & \checkmark & & GPT(3.5, 4), Claude, Gemini Advanced\\ \hline
2024 &~\cite{vitolo2024natural} & \checkmark & & \checkmark & & GPT-4\\ \hline
2024 &~\cite{xiang2024digital} & & & \checkmark & \checkmark & ChatGPT-4\\ \hline
2024 &~\cite{10691811} & \checkmark & & \checkmark & & ChatGPT(3.5, 4), Bard, HuggingChat\\ \hline
2024 & RTLCoder~\cite{liu2024rtlcoder} & \checkmark & & & & RTLCoder based on Mistral-7B and DeepSeek-Coder-6.7B\\ \hline
2024 &~\cite{nadimi2024multiexpert} & \checkmark & & & & CodeGen(2B, 6B, 16B), GEMMA(2B, 7B)\\ \hline
2024 &~\cite{10462177} & & & & \checkmark & GPT-4, Codex, CodeGen\\ \hline
2024 &~\cite{10.1109/ASP-DAC58780.2024.10473927} & & & \checkmark & & Hardware Phi-1.5B\\ \hline
2024 &~\cite{meech2024leveraginghighlevelsynthesislarge} & \checkmark & \checkmark & \checkmark & & Microsoft Bing Chat\\ \hline
2024 &~\cite{10.1007/978-3-031-63378-2_72} & \checkmark & & & & Llama2-70B\\ \hline
2024 &~\cite{chang2024naturallanguageenoughbenchmarking} & & & \checkmark & & GPT(4, 4V), LLaVA, LLaMA\\ \hline
2024 &~\cite{zhao2024codevempoweringllmsverilog} & & & \checkmark & \checkmark & GPT-3.5, CodeLlama, DeepSeekCoder, CodeQwen\\ \hline
2024 &~\cite{jiang2024iicpilotintelligentintegratedcircuit} & \checkmark & \checkmark & & & GPT(3.5, 4)\\ \hline
2024 &~\cite{abdelatty2024hdlcopilothardwaredesignlibrary} & & & \checkmark & & GPT(3.5, 4, 4o)\\ \hline
2024 &~\cite{he2024chatedalargelanguagemodel} & \checkmark & & & & AutoMage fine-tuned on Llama2-70B, GPT(3.5, 4)\\ \hline
2024 &~\cite{ho2024largelanguagemodelllm} & & \checkmark & & & GPT-3.5-turbo-16k-0613\\ \hline
2024 &~\cite{gao2024autovcodersystematicframeworkautomated} & \checkmark & & \checkmark & & Codellama-7B, DeepSeek-Coder-6.7B, CodeQwen1.5-7B\\ \hline
2024 &~\cite{nakkab2024romebuiltsinglestep} & \checkmark & & & & Llama 2, Code Llama, VeriGen, CL-Verilog, RTL-Coder, Llama 3, GPT(3.5, 4)\\ \hline

2025 & Opl4gpt~\cite{tasnia2025opl4gpt} & \checkmark & & & & GPT-4\\ \hline
2025 &~\cite{zhao2025enhancing} & \checkmark & & & & DeepSeek-Coder-6.7B, RTLCoder-DeepSeek-v1.1\\ \hline
2025 & Chipmnd~\cite{firouzi2025chipmnd} & \checkmark & & \checkmark & & GPT(3.5, 4), DeepSeek R1, Claude 3.5, Llama 3\\ \hline
2025 & Fixbench-RTL~\cite{li2025fixbench} & & & & \checkmark & GPT, Claude, Gemini, DeepSeek, Qwen, and Llama\\ \hline
2025 &~\cite{elnaggar2025adding} & & & & \checkmark & GPT(4, 4o)\\ \hline
2025 &~\cite{qayyum2025llm} & & & & \checkmark & GPT(4, 4 Turbo)\\ \hline
2025 & FLAG~\cite{ahmad2025flag} & & & \checkmark & \checkmark & GPT-3.5 Turbo\\ \hline
2025 & VeriReason~\cite{wang2025verireason} & \checkmark & & & & CodeLlama/Qwen + GRPO RL\\ \hline
2025 & VeriRL~\cite{teng2025verirl} & \checkmark & \checkmark & & & LLMs + RL, Veribench-53K\\ \hline
2025 & VeriCoder~\cite{wei2025vericoder} & \checkmark & & \checkmark & & GPT-4o-mini for data generation\\ \hline
2025 & SAGE-HLS~\cite{khan2025sagehls} & \checkmark & \checkmark & & & QwenCoder-7B, AST-guided\\ \hline
2025 & HLStrans~\cite{zou2025hlstrans} & & \checkmark & & & LLMs, C-to-HLS dataset\\ \hline
2025 & RocketPPA~\cite{abdollahi2025rocketppa} & & \checkmark & & & LLM + Mixture-of-Experts\\ \hline
2025 & CRADLE~\cite{krupp2025cradle} & & \checkmark & \checkmark & & o4-mini, multi-agent\\ \hline
2025 & EDAid~\cite{wu2025edaid} & \checkmark & \checkmark & & & ChipLlama, multi-agent\\ \hline
2025 & AutoEDA~\cite{lu2025autoeda} & & \checkmark & & & LLM agents, RTL-to-GDSII\\ \hline
2025 & MACO~\cite{jiang2025maco} & \checkmark & \checkmark & \checkmark & & LLM agents, CGRA HW/SW\\ \hline
2025 & VeriDebug~\cite{wang2025veridebug} & & & & \checkmark & Unified LLM, contrastive\\ \hline
2025 &~\cite{firouzi2025prompt} & \checkmark & \checkmark & & & GPT-4\\ \hline

\end{tabularx}
\end{table*}
Overall, debugging-oriented LLM research shows a progression from simple syntax correction toward retrieval-enhanced debugging, diagnostic explanation, feedback-driven repair, and security-aware bug fixing. The strongest approaches do not rely on the LLM alone; instead, they combine language-model reasoning with compiler diagnostics, simulator feedback, retrieved documentation, or external validation. This suggests that LLMs are most useful for hardware debugging when integrated into feedback-driven EDA workflows rather than used as standalone code-repair tools.

Table~\ref{LLM_design} provides a comprehensive view of different methods and AI models used in digital hardware design from 2022 to 2025. It further classifies these methods into the year they were introduced and their capabilities regarding design assistance, optimization, synthesis, verification, validation, debugging, and maintenance. The last column lists the associated AI models used in each paper, such as GPT, CodeGen, etc. This table highlights the evolution and integration of AI-driven tools in digital hardware design, focusing on methods that involve various levels of assistance in the design and verification process. The presence of diverse models, such as GPT-3.5, GPT-4, and versions of Llama, and some fine-tuned models, such as VeriGen and CodeGen, indicates a rise in the use of powerful AI models over time to support hardware-related tasks. 

\subsection{LLM-based Analog Hardware Design}  
LLMs have demonstrated transformative potential not only in digital systems but also in analog hardware design, where they are gaining traction for topology generation, circuit sizing, simulation-guided optimization, and layout assistance. Unlike digital design, analog design involves continuous parameter tuning, strict physical constraints, and close interaction between schematic-level behavior and layout-level implementation. As a result, LLM-based analog design frameworks often combine language-model reasoning with simulation feedback, optimization algorithms, reusable circuit libraries, or multi-agent layout workflows.
A key area of focus is using LLMs to automate and assist in analog circuit design. Lai et al. introduced AnalogCoder, a training-free LLM agent designed specifically for analog circuit design through Python code generation~\cite{lai2024analogcoderanalogcircuitdesign}. AnalogCoder leverages a feedback-enhanced design flow and a comprehensive circuit tool library to automate the design process. It successfully designed 20 out of 24 benchmark circuits, outperforming other LLM-based methods. Its feedback mechanisms, including requirement checks, simulation and operating-point checks, DC sweep checks, and function checks, significantly improved design success rates. By archiving successful designs as reusable, modular sub-circuits, AnalogCoder also simplified the creation of more complex circuits.

Similarly, Chang et al. introduced LaMAGIC, a framework that uses language models to generate optimized analog circuit topologies from user-defined specifications, particularly for power converter applications~\cite{chang2024lamagiclanguagemodelbasedtopologygeneration}. Leveraging supervised fine-tuning, LaMAGIC can produce high-quality circuit designs in a single pass, significantly reducing design time compared to traditional methods. The framework demonstrated a success rate of up to 96\% under strict tolerances and showed superior performance with adjacency-matrix-based formulations for more complex circuits. Together, AnalogCoder and LaMAGIC show how LLMs can support early-stage analog design through specification interpretation, topology generation, feedback-guided refinement, and reusable design knowledge.
Another significant application of LLMs is in optimizing analog circuit design processes. Chen et al. presented LLANA, a framework that leverages LLMs to enhance Bayesian optimization for generating analog design-dependent parameter constraints~\cite{chen2024llmenhancedbayesianoptimizationefficient}. LLANA reduces simulation requirements while maintaining adaptability across different circuit designs, achieving results comparable to state-of-the-art Bayesian optimization methods with fewer simulations. The framework's success was demonstrated in the design of two-stage operational amplifiers, where it achieved better optimization with fewer samples.

Similarly, Yin et al. introduced ADO-LLM, a framework that combines Bayesian optimization with LLMs to improve analog circuit design efficiency and effectiveness~\cite{yin2024adollmanalogdesignbayesian}. ADO-LLM leverages LLMs' domain knowledge to generate initial design points and uses Gaussian process-based Bayesian optimization to optimize these points iteratively. The framework demonstrated notable improvements in design efficiency and performance on two analog circuits: a two-stage differential amplifier and a hysteresis comparator.

\begin{figure*}[htbp]     
    \centering
    \includegraphics[width=0.95\linewidth]{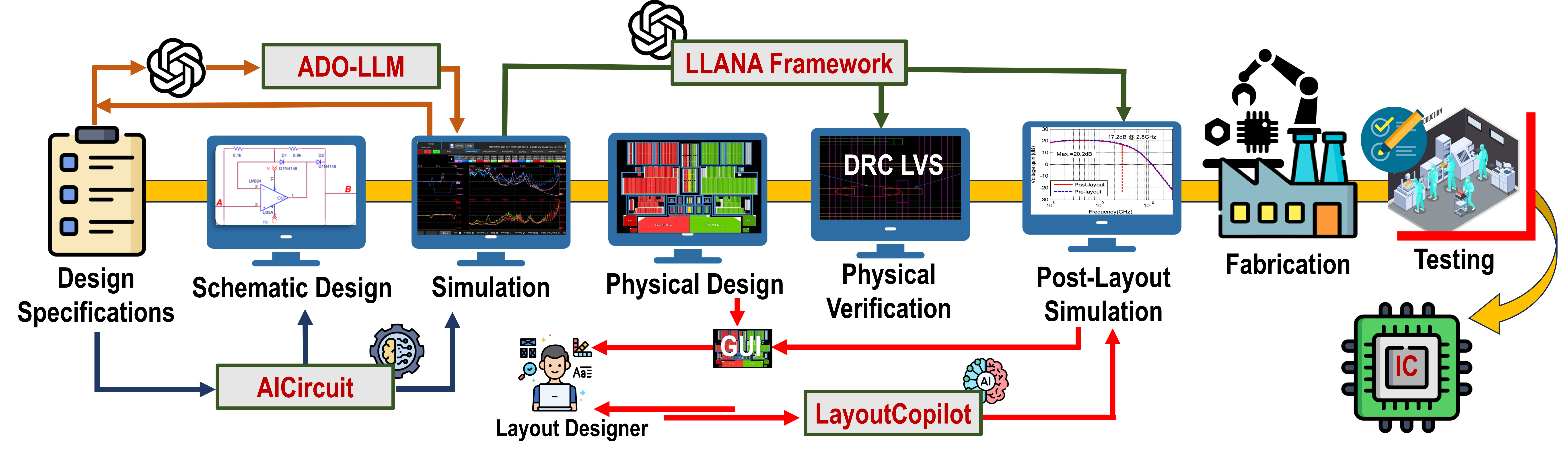}
    \caption{Key stages in the IC design flow showcasing the integration of AI frameworks like ADO-LLM, LLANA, AICircuit, and LayoutCopilot for optimizing schematic, simulation, and physical design processes.}
    \label{fig:analogdesign}
\end{figure*}

Both LLANA and ADO-LLM highlight the synergy between LLMs and optimization algorithms, where language models provide domain-informed initialization or constraint guidance while numerical optimization performs detailed parameter search. This combination is particularly useful in analog design because performance depends on continuous design variables, simulation feedback, and trade-offs among gain, bandwidth, power, stability, and layout constraints.

LLMs have also been applied to analog layout design, where they assist in optimizing physical layouts and improving verification processes. Chang et al. introduced LayoutCopilot, an LLM-powered multi-agent collaborative framework designed to enhance the interactive analog layout design process~\cite{liu2024layoutcopilotllmpoweredmultiagentcollaborative}. This framework integrates LLMs within a multi-agent system to streamline and optimize layout tasks, achieving 6.49\% improvement in logical verification, 4.85\% improvement in syntactic verification, and 5.89\% improvement in overall accuracy using structured instructions. Multi-agent configurations proved more effective, achieving a correctness rate of 96.80\% with Claude-3. After interactive adjustment, the layout area was reduced to 66\% of its original size, with additional gains in performance metrics such as gain, phase margin, and common-mode rejection ratio.


The 2025 literature pushes analog LLM assistance toward foundation models and reinforcement learning. Chen et al. released AnalogSeeker, an open-source foundation language model for analog circuit design trained on a curated analog-circuit corpus with granular knowledge distillation; it attains 85.04\% accuracy on the AMSBench-TQA knowledge benchmark and demonstrates downstream operational-amplifier design ability while remaining fully open~\cite{chen2025analogseeker}. Vijayaraghavan et al. proposed AutoCircuit-RL, which combines instruction tuning with reinforcement learning to synthesize analog topologies directly from specifications, generating roughly 12\% more valid circuits with about 14\% better efficiency than prior methods even under limited training data~\cite{vijayaraghavan2025autocircuitrl}. These works move analog design beyond per-task prompting toward reusable, domain-adapted models and feedback-driven topology generation.

\begin{table*}[!t]
\centering
\caption{LLM-Based Analog Hardware Design}
\label{tab:llm_based_analog_design}

\renewcommand{\arraystretch}{1.15}
\setlength{\tabcolsep}{5pt}
\small

\begin{tabularx}{\textwidth}{|
>{\centering\arraybackslash}m{0.9cm}|
>{\raggedright\arraybackslash}X|
>{\raggedright\arraybackslash}m{3.6cm}|
>{\raggedright\arraybackslash}X|}
\hline

\rowcolor[HTML]{E3E3E3}
\textbf{Year} &
\textbf{Method Name} &
\textbf{Application} &
\textbf{AI Model} \\
\hline

2024 &
LLANA~\cite{chen2024llmenhancedbayesianoptimizationefficient} &
Design Assistance &
GPT-3.5 \\
\hline

2024 &
LaMAGIC~\cite{chang2024lamagiclanguagemodelbasedtopologygeneration} &
Design Assistance &
Language Model \\
\hline

2024 &
Analog Coder~\cite{lai2024analogcoderanalogcircuitdesign} &
Design Assistance &
GPT-3.5 \\
\hline

2024 &
ADO-LLM~\cite{yin2024adollmanalogdesignbayesian} &
Optimization &
GPT-3.5 Turbo \\
\hline

2024 &
Layout Copilot~\cite{liu2024layoutcopilotllmpoweredmultiagentcollaborative} &
Layout Design &
GPT-3.5, GPT-4, Claude 3 \\
\hline

2025 &
AnalogSeeker~\cite{chen2025analogseeker} &
Foundation Model / Design Assistance &
Qwen2.5-32B \\
\hline

2025 &
AutoCircuit-RL~\cite{vijayaraghavan2025autocircuitrl} &
Topology Generation &
LLM + Reinforcement Learning \\
\hline

\end{tabularx}
\end{table*}

 



    

Figure ~\ref{fig:analogdesign} illustrates the integration of AI frameworks into the analog IC design flow. The figure primarily highlights how these frameworks optimize various IC design stages. All these AI frameworks are utilized to enhance the efficiency and precision of the hardware design process. This figure showcases the underlying roles of these AI tools in streamlining the analog IC design workflow.
Table ~\ref{tab:llm_based_analog_design} summarizes recent works that apply LLMs to analog hardware design. This table lists significant papers from 2024 to 2025, highlighting their applications and the AI models they use. The papers: LLANA, LaMAGIC, and Analog Coder focus specifically on design assistance, whereas ADO-LLM and Layout Copilot address optimization and layout design, respectively. In these studies, researchers employ advanced LLMs such as GPT-3.5, GPT-3.5 Turbo, GPT-4, and Claude 3, showcasing their ability to handle complex design processes. These papers emphasize how LLMs increasingly enhance efficiency and accuracy throughout the analog hardware design workflow.

\subsection{Other Attention-based Hardware Design}

\begin{figure}[t]
    \centering
    \includegraphics[width=1\linewidth]{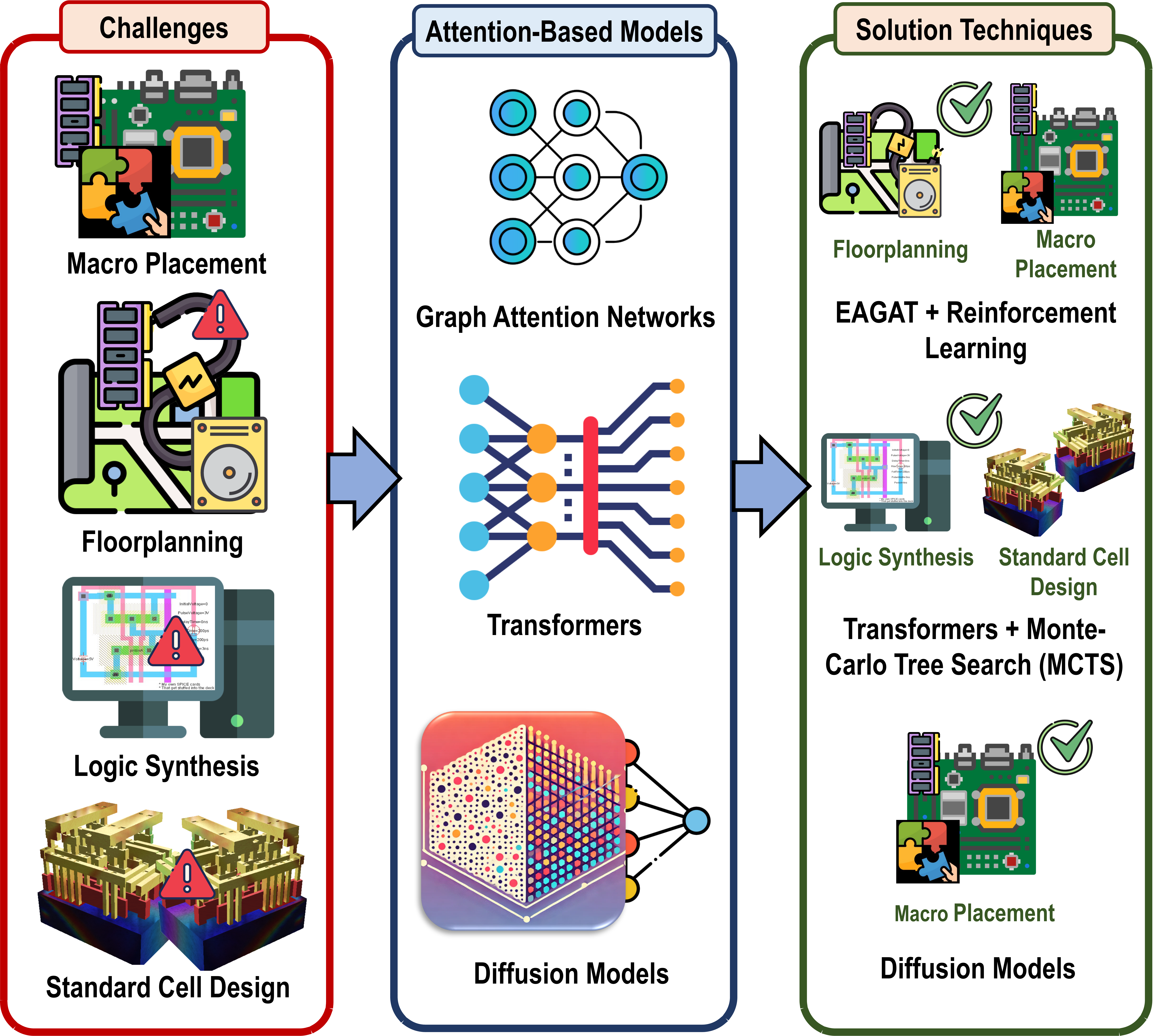}
    \caption{Overview of attention-based hardware design techniques leveraging models like graph attention networks (GATs), transformers, and diffusion models to address challenges in hardware design.} 
    \label{fig:attentionbased}
\end{figure}

Beyond LLM-based design assistance, attention-based models have also been applied directly to structured hardware-design representations such as circuit graphs, placement states, layout images, and synthesis trajectories. These methods are distinct from LLM-centric approaches because their primary interface is not natural-language or HDL token generation, but structured learning over hardware artifacts. As shown in Figure~\ref{fig:attentionbased}, attention-based models have been explored for floorplanning, macro placement, logic synthesis, standard-cell design, circuit representation learning, and hotspot detection.

One prominent application area is floorplanning and macro placement, where optimizing chip area, wire length, placement legality, and runtime remains challenging. Yang et al. introduced an end-to-end reinforcement learning methodology for chip floorplanning that optimizes macro placement and orientation to minimize chip area and wire length using an edge-aware graph attention network (EAGAT)~\cite{yang2024floorplanning}. Similarly, Lee et al. proposed a diffusion-model-based approach for macro placement to address limitations of sequential reinforcement learning methods~\cite{lee2024chipplacementdiffusion}. Their method places all macros simultaneously and uses a neural architecture combining graph convolutions and multi-head attention layers for efficient and expressive placement. The authors also introduced a strategy for generating large synthetic datasets for pre-training, reducing dependence on proprietary placement data. Experimental results showed competitive performance on benchmark datasets in terms of legality and half-perimeter wire length (HPWL).

Transferability and automation are also important in chip placement, where learned policies must generalize to new designs while reducing runtime. Here, transferability refers to the ability of a learned placement policy or optimization strategy to generalize from training circuits to previously unseen chip designs without requiring full retraining from scratch. Lai et al. explored this direction with ChiPFormer, a transformer-based placement framework that uses offline reinforcement learning to learn a transferable placement policy~\cite{lai2023chipformertransferablechipplacement}. ChiPFormer reduced placement runtime by up to 10$\times$ and achieved superior HPWL compared to recent reinforcement-learning-based methods. The framework transferred learned policies to new chip designs, reducing placement time from hours to minutes and outperforming methods such as GraphPlace, MaskPlace, and DeepPR across 32 chip circuits.

In logic synthesis and standard-cell design, attention-based models are useful because they can capture graph-structured dependencies and long-range design relationships. Li et al. introduced Circuit Transformer, a transformer-based model for end-to-end logic synthesis that predicts the next logic gate during circuit construction~\cite{li2024circuittransformerendtoendcircuit}. The framework uses a novel encoding scheme and integrates Monte Carlo Tree Search to improve generated circuit designs. By combining transformer-based prediction with search-based refinement, Circuit Transformer demonstrates how attention models can support synthesis tasks that require both structural understanding and sequential decision-making.

In standard-cell design, Ho et al. introduced a transformer-based clustering methodology to improve standard-cell layout automation~\cite{10.1145/3626184.3633314}. By training the model with LVS/DRC-clean cell layouts and using personalized PageRank vectors, the authors addressed routability challenges in advanced nodes. Their method generated 15\% more LVS/DRC-clean layouts, reduced average cell width by 3.9\%, reduced total wire length by 3.3\%, and achieved a 12.7$\times$ speedup over previous methods. The study also demonstrated scalability by achieving 100\% LVS/DRC-clean layouts for more than 1000 single-row cells and reducing cell area by 14.5\% compared to industrial standards.

Scalability and generalizability remain key challenges when applying attention-based models to large-scale circuit design. Deng et al. introduced Hop-Wise Graph Attention (HOGA), a method designed to improve the scalability and generalizability of GNNs for circuit representation learning~\cite{deng2024morehopwisegraphattention}. HOGA precomputes hop-wise features to support efficient distributed training and adaptively captures high-order circuit structures. Evaluations on the OpenABC-D benchmark and functional reasoning tasks showed that HOGA reduced Quality-of-Result (QoR) estimation error by 46.76\% and improved reasoning accuracy by 10\% compared to conventional GNNs. Although HOGA introduces additional computational cost, its architecture supports massive parallelization and is suitable for large-scale circuit datasets.

Attention-based models have also been applied to hotspot detection in semiconductor layout design, which is critical for design for manufacturability. Zhu et al. introduced a single-stage end-to-end hotspot detection framework that uses multi-task learning and a transformer encoder to improve hotspot detection efficiency and accuracy~\cite{9643590}. The framework uses center and corner detection heads for representation learning and a transformer encoder for global feature aggregation. Experiments on the ICCAD CAD Contest 2016 benchmarks showed 97.31\% detection accuracy, up to 67.84$\times$ speedup over prior methods, and substantial false-alarm reduction compared with earlier TCAD'19 and DAC'19 approaches.

Researchers have also explored attention-based models for analog and radio-frequency circuit design. Mehradfar et al. introduced AICircuit, a multi-level dataset and benchmark for advancing ML-based analog and RF circuit design~\cite{mehradfar2024aicircuitmultileveldatasetbenchmark}. AICircuit includes seven fundamental circuits and two complex wireless transceiver systems, providing a resource for evaluating ML models that map design specifications to circuit parameters. The authors demonstrated the potential of models such as multilayer perceptrons and transformers for simplifying analog design, while also showing that highly nonlinear and complex circuits remain challenging for current ML methods.


Attention- and graph-based learning for physical design also advanced in 2025. Hou et al. proposed TransPlace, a graph-neural-network framework for transferable circuit global placement that learns reusable placement knowledge and applies it to unseen circuits, achieving a 1.2$\times$ speedup while reducing congestion by 30\%, timing violations by 9\%, and wirelength by 5\%~\cite{hou2025transplace}. This continues the trend, observed throughout this section, of attention- and graph-based models being used not as text generators but as structured optimizers that exploit circuit topology and transfer across designs.

Table~\ref{tab:attention_based_hardware_design} summarizes representative works using attention-based models for hardware design. The table covers applications including hotspot detection, static timing analysis, placement, floorplanning, logic synthesis, standard-cell design automation, and circuit representation learning. These works use a range of attention-based architectures, including multi-head transformers, GAT-based models, decision transformers, edge-aware graph attention networks, and hop-wise graph attention. Together, they show that attention mechanisms are increasingly being used not only for text-oriented design assistance, but also for structured hardware optimization and representation learning.
\begin{table*}[!t]
\centering
\caption{Other Attention-Based Hardware Design}
\label{tab:attention_based_hardware_design}

\renewcommand{\arraystretch}{1.25}
\setlength{\tabcolsep}{4pt}

\begin{tabularx}{\textwidth}{|
>{\centering\arraybackslash}m{1.1cm}|
>{\raggedright\arraybackslash}X|
>{\raggedright\arraybackslash}X|
>{\raggedright\arraybackslash}X|}
\hline

\rowcolor[HTML]{E3E3E3}
\textbf{Year} &
\textbf{Method Name} &
\textbf{Application} &
\textbf{AI Model} \\
\hline

2021 &
Single-stage detector~\cite{9643590} &
Hotspot Detection &
Multi-headed transformer \\
\hline

2023 &
Deep EdgeGAT~\cite{ye2023graph} &
STA Analysis &
GAT-based model \\
\hline

2023 &
ChiPFormer~\cite{lai2023chipformertransferablechipplacement} &
Placement &
Decision Transformer \\
\hline

2024 &
Hierarchical Floorplanning Network (HFN)~\cite{yang2024floorplanning} &
Floorplanning &
Edge-aware graph attention network \\
\hline

2024 &
Circuit Transformer~\cite{li2024circuittransformerendtoendcircuit} &
Logic Synthesis &
Transformer-based model \\
\hline

2024 &
Transformer-Based Clustering Model~\cite{10.1145/3626184.3633314} &
Standard Cell Design Automation &
Transformer model \\
\hline

2024 &
HOGA~\cite{deng2024morehopwisegraphattention} &
Representation &
Attention-based model \\
\hline

2025 &
TransPlace~\cite{hou2025transplace} &
Global Placement &
GNN-based transferable model \\
\hline

\end{tabularx}
\end{table*}

Overall, these works show that non-LLM attention-based methods complement LLM-centric approaches by operating directly on structured hardware representations such as circuit graphs, placement states, layouts, and synthesis trajectories. Their main strength lies in modeling topology, long-range design dependencies, and optimization context, especially when combined with reinforcement learning, search, or domain-specific EDA constraints.

\section{Attention-Based hardware security}
\label{sec:AttentionBasedHardwareSecurity}

\subsection{LLM-based Hardware Security}

This section presents the role of LLMs in enhancing hardware security on various tiers of applications, such as vulnerability detection, threat analysis, security verification, and logic obfuscation. We also explore frameworks on automated debugging, security assertion generation, and thwarting side-channel attacks. A common theme running across these works is to address the challenge of scalability, accuracy, and efficiency in hardware security using LLMs.

\begin{figure}[!htbp]
    \centering
    \includegraphics[width=1\linewidth]{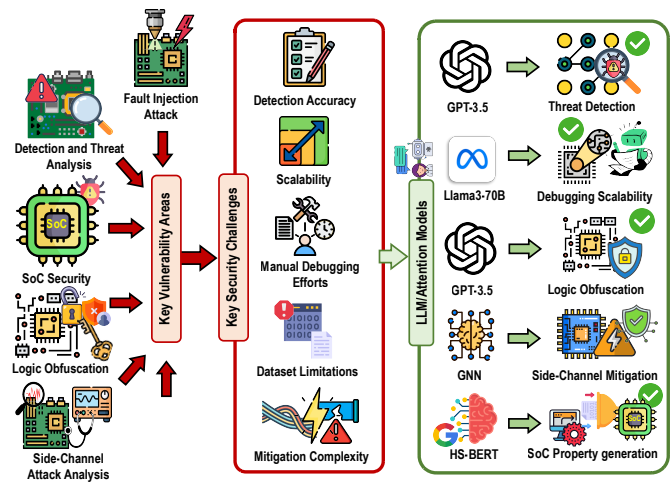}
    \caption{Overview of security vulnerabilities, challenges, and the role of LLMs and Attention-based models in detection, mitigation, and debugging for hardware security.} 
    \label{fig:security}
\end{figure}

Figure~\ref{fig:security} highlights key security vulnerabilities and challenges in hardware design, including fault injection attacks, SoC security, logic obfuscation, and side-channel attacks. It illustrates the potential of LLM/attention models such as GPT-3.5, Llama3-70B, GNNs, and HS-BERT in enhancing security by addressing detection accuracy, scalability, debugging efforts, and dataset limitations, enabling effective threat detection, logic obfuscation, side-channel mitigation, and SoC property generation.


A key dimension in comparing hardware-security methods is the design stage at which they operate. Some approaches analyze specifications, documentation, or security policies at the system or SoC level; others target RTL or SystemVerilog artifacts for vulnerability detection, assertion generation, bug repair, or Trojan insertion; still others operate on structural graph representations or post-design security signals such as side-channel traces. Because each abstraction level exposes different information and constraints, results are not always directly comparable across papers. In the discussion below, we therefore distinguish methods according to whether they act primarily on documentation and specification artifacts, RTL and HDL representations, structural graph-level representations, or post-design attack signals~\cite{gadde2024artificialintelligencegenailens},~\cite{kande2023llmassisted},~\cite{meng2023unlocking},~\cite{10508974},~\cite{tarek2024socurellm}.

\subsubsection{Vulnerability Detection And Threat Analysis}
A significant focus of recent research has been on leveraging LLMs for vulnerability detection and threat analysis in hardware systems. 
Ghimire et al. presented the HWREx framework, an NLP-based system focused on IoT hardware vulnerabilities~\cite{Ghimire2025_Hwrex}. This framework uses ontology-driven storytelling to analyze and mitigate vulnerabilities, predict future exposures, and provide mitigation suggestions using GPT models. The framework offers an interactive GUI for identifying mitigation strategies and determining exploit and impact scores with the Common Vulnerability Scoring System. Similarly, Saha et al. presented Vul-FSM, a database containing 10,000 vulnerable finite state machine (FSM) designs featuring 16 specific security vulnerabilities~\cite{10545393}. The paper introduces SecRT-LLM, leveraging LLMs to create extensive hardware security datasets efficiently. Results demonstrate the framework's effectiveness: initial attempts achieve an average pass rate of 81.98\% for vulnerability insertion and 97.37\% for detection, improving to 80.30\% and 99.07\%, respectively, within five attempts. This highlights LLMs' proficiency in enhancing ML-based methods for hardware security benchmarking and mitigation strategies.

Another critical area of research is the automation of security verification and debugging processes. Akyash et al. introduced Self-HWDebug, a framework designed to enhance the scalability and efficiency of LLMs in automating the mitigation of security vulnerabilities in hardware designs~\cite{akyash2024selfhwdebug}. The framework leveraged LLMs to autonomously generate detailed debugging instructions, addressing the challenge of manually crafting instructions that demand significant time and expertise. Self-HWDebug prompts LLMs to create targeted mitigation instructions by utilizing pre-identified hardware bugs and resolutions, extending solutions across similar vulnerabilities in different designs. Initial testing shows the framework's efficacy in reducing human intervention and improving debugging quality. 
Complementing this work, Saha et al. explored the integration of LLMs, particularly GPTs, into the security verification process of SoC designs~\cite{saha2023llm}. Current security solutions struggle with scalability and adaptability, making them inadequate for modern SoCs. The research aims to create a more efficient and comprehensive security verification methodology by leveraging LLMs. The paper analyzes existing works, provides case studies and experiments, and offers guidelines for using LLMs in SoC security. It highlights the potential benefits and challenges of this innovative approach. 

Formal verification has emerged as a key technique for identifying and mitigating Common Weakness Enumerations (CWEs) in LLM-generated hardware designs. Gadde et al. introduced ReFormAI, a dataset of 60,000 SystemVerilog designs generated by various LLMs, focusing on identifying and addressing CWEs through formal verification, demonstrating that approximately 60\% of LLM-generated designs are prone to CWEs, and highlighting the superior reliability of formal verification over traditional testing methods. Gadde et al. investigated the prevalence of CWEs in SystemVerilog hardware designs generated by LLMs and evaluated the effectiveness of formal verification in identifying these vulnerabilities~\cite{gadde2024artificialintelligencegenailens}. They introduced the ReFormAI dataset, comprising 60,000 SystemVerilog designs generated by four different LLMs targeting various CWEs. The study revealed a high vulnerability rate, with approximately 60\% of the generated designs containing CWEs. Formal verification techniques successfully identified and categorized these vulnerabilities. Among the evaluated LLMs, GPT-3.5-Turbo performed best but still produced a significant number of vulnerable designs. The study underscores the need for better training datasets and fine-tuning techniques to reduce CWEs in LLM-generated hardware designs.
 Building on this, Fu et al. presented Hardware Phi-1.5B, an LLM tailored explicitly for the hardware domain, addressing challenges in hardware design and security verification~\cite{10.1109/ASP-DAC58780.2024.10473927}. The model leveraged a specialized, tiered dataset focused on hardware-specific content, demonstrating advancements in predicting the next tokens and handling complex hardware design tasks. Experimental results validated the model's proficiency, particularly on the Common Weakness Enumeration (CWE) dataset.

 The automation of hardware security assertion generation is another promising application of LLMs. Kande et al. investigated the potential of LLMs in automating the generation of hardware security assertions~\cite{kande2023llmassisted}. By introducing a new evaluation framework and benchmark suite, the study explored the effectiveness of LLMs in generating correct security assertions for hardware designs. OpenAI's Codex showed a modest success rate, generating correct assertions 4.53\% of the time, with detailed prompts improving performance. The study used Siemens Modelsim for simulation and validation, demonstrating the framework's scalability to other LLMs. Despite challenges, such as the need for detailed prompts and context, the results provide a proof-of-concept for LLM-assisted assertion generation, suggesting future work on fine-tuning models, improving datasets, and exploring quick evaluation methods to enhance accuracy and effectiveness in hardware security applications.

Recent efforts have also explored the use of LLMs for automated Trojan detection and mitigation in analog designs. Chaudhuri et al. developed SPICED+, a software-based framework that leverages LLM-enhanced analysis of SPICE netlists and simulation logs to iteratively detect, localize, and remove analog hardware Trojans~\cite{chaudhuri2025spiced}. The framework combines rule-based prompting, few-shot learning, and voltage/current deviation analysis to classify Trojan-impacted nodes and applies an iterative correction loop guided by a confidence scoring mechanism. By integrating with HSPICE for netlist resimulation and validation, SPICED+ achieves an average Trojan coverage of 93.3\%, a mitigation rate of 94.6\%, and maintains a low false positive rate of 1.4\%, outperforming prior ML-based tools like DAWN. Its model-agnostic LLM workflow enables generalization to new circuits without retraining, though large-scale designs such as SAR-ADCs may require modular decomposition for efficient context processing.

The 2025 literature on vulnerability detection is increasingly defined by fine-tuned open models and multi-agent reasoning. Tarek et al. introduced BugWhisperer, which applies parameter-efficient Low-Rank Adaptation (LoRA) to fine-tune open-source LLMs for SoC vulnerability detection on a purpose-built hardware-vulnerability database, substantially outperforming both non-fine-tuned and general-purpose counterparts and showing that tailored open models can rival proprietary ones on specialized security tasks~\cite{tarek2025bugwhisperer}. Collini et al. proposed MARVEL, a multi-agent framework in which supervisor and executor agents combine decision-making, tool use, and reasoning to extract RTL vulnerabilities, surfacing 19 valid vulnerabilities in a buggy OpenTitan SoC~\cite{collini2025marvel}. Ahmad et al. paired LLMs with static analysis in LASHED, using the model to prune false positives and explain security impact across four open-source SoCs and five CWEs, with 87.5\% of flagged instances judged plausible vulnerabilities~\cite{ahmad2025lashed}. Long et al. took a rule-generation route with VerilogLAVD, introducing a Verilog Property Graph that unifies AST syntax with control- and data-flow semantics so that LLMs can synthesize CWE-derived detection rules, raising F1 to 0.54 across 77 designs and 12 CWE types~\cite{long2025veriloglavd}. These efforts share a common lesson with the rest of this survey: detection accuracy improves most when LLMs are grounded in structured representations, curated datasets, or external analysis engines.
 
\subsubsection{SoC Security}
We treat SoC security as a distinct category because it is fundamentally a system-level security problem rather than a purely module-local RTL problem. In modern SoCs, security depends not only on the correctness of individual blocks, but also on the interactions among heterogeneous IPs through buses, memory hierarchies, privilege boundaries, debug interfaces, and firmware-visible control paths. This makes SoC security especially dependent on documentation analysis, policy extraction, property generation, and cross-module reasoning. By contrast, other security topics discussed in this survey, such as Trojan insertion, localized RTL vulnerability repair, and some forms of logic obfuscation, often focus more directly on specific design artifacts or lower-level representations. Separating SoC security from these topics, therefore, reflects a difference in abstraction level and verification objective, rather than a claim that SoC threats are unrelated to the broader hardware-security landscape~\cite{meng2023unlocking,saha2023llm,tarek2024socurellm}.

LLM-based tools have demonstrated significant potential in automating and enhancing the security verification process for SoC designs. These tools address the limitations of traditional methods by leveraging NLP and LLMs to extract security properties, identify vulnerabilities, and generate comprehensive security policies. For instance, Meng et al introduced NSPG, an innovative NLP-based tool designed to automate the generation of security properties from SoC documentation~\cite{meng2023unlocking}. The tool leverages the HS-BERT model, a hardware security-specific language model trained on sentences from RISC-V, OpenRISC, MIPS, OpenSPARC, and OpenTitan SoC documentation. In evaluations on five untrained OpenTitan hardware IP documents, NSPG successfully extracted 326 security properties from 1723 sentences, aiding in the identification of eight security bugs presented in the Hack@DAC 2022 competition. This approach demonstrates superior performance compared to traditional methods and popular tools like ChatGPT, highlighting its potential to streamline the security validation process and improve the robustness of SoC designs. The success of NSPG suggests a promising future for NLP applications in hardware security and SoC verification. Similarly, Tarek et al. introduce SoCureLLM, an LLM-based framework designed to improve hardware security verification for complex SoC designs~\cite{tarek2024socurellm}. SoCureLLM addresses the limitations of traditional methods by effectively identifying security vulnerabilities and generating a comprehensive security policy database. In evaluations, it detected 76.47\% of security bugs in three vulnerable RISC-V SoCs, surpassing existing methods. Additionally, SoCureLLM formulated 84 new security policies for large-scale RISC-V SoC designs, enhancing the security policy database. This framework demonstrates significant improvements in adaptability, scalability, Trojan insertion analysis, and efficiency for SoC security verification.


Two 2025 frameworks advance the system-level, documentation-driven view of SoC security emphasized in this subsection. Ankireddy et al. proposed LASA, which couples an LLM with retrieval-augmented generation to produce non-vacuous security properties and SystemVerilog assertions from SoC specifications, integrating a commercial formal-property-verification tool and iteratively refining prompts via coverage feedback to reach roughly 88\% coverage while uncovering five unique bugs in the Hack@DAC'24 OpenTitan SoC~\cite{ankireddy2025lasa}. Saha et al. introduced SV-LLM, an agentic system that decomposes SoC security verification into specialized agents for security-asset identification, threat modeling, test-plan and property generation, vulnerability detection, and simulation-based bug validation, combining in-context learning, fine-tuning, and retrieval into an end-to-end workflow~\cite{saha2025svllm}. Both works reinforce the argument that SoC security is fundamentally a cross-module, specification-level problem that benefits from coordinated, tool-grounded LLM reasoning rather than isolated module analysis.


\subsubsection{Trojan Insertion}
Research in hardware security has increasingly leveraged LLMs for generating, inserting, and detecting hardware trojans. These studies demonstrate the dual-use nature of LLMs, addressing challenges such as dataset scarcity, context length limitations, and the need for diverse HT examples in hardware design.
Kokolakis et al. investigate the potential of LLMs in offensive hardware security, particularly for inserting hardware trojans into complex hardware designs like CPUs~\cite{kokolakis2024harnessing}. They test how well LLMs can correlate high-level security concepts with specific hardware modules to overcome context length limitations. By analyzing reduced code bases and identifying modules with security-related features, they simplify the overall analysis. They demonstrate the LLM’s capability to modify code and insert HT functionalities. The approach is validated by creating and testing an HT in a RISC-V micro-architecture on an FPGA board, showcasing the LLM's ability to aid in realistic HT attacks.

Complementing this work, Hayashi et al. developed a comprehensive dataset consisting of 290 Verilog examples generated using ChatGPT-4. This dataset~\cite{hayashi2024hardware}, derived from 29 golden models and the TrustHub taxonomy, significantly enhances the limited existing datasets for hardware trojans, particularly for RISC-V and Web3 applications. It addresses the gap in research on Web3 hardware trojans, including modules like hardware wallets. This new dataset is intended to facilitate future research on defense mechanisms against hardware trojans in RISC-V systems, hardware wallets, and hardware Proof of Work (PoW) miners.
Building on these advancements, Bhandari et al. introduced SENTAUR, a framework leveraging to generate, detect, and assess hardware Trojans in RTL designs~\cite{bhandari2024sentaursecurityenhancedtrojan}, addressing security challenges in the IC supply chain. SENTAUR utilizes GPT-4 to create synthesizable RTL with embedded HTs, validated on the Xilinx FPGA platform. It explores various HT triggers and effects, demonstrating effectiveness in real-world applications like AES-T800 and dual-port RAM. A comparative analysis showed SENTAUR's superiority over existing tools in generating diverse and effective hypotheses. Experimental results highlighted SENTAUR's efficiency and robustness, demonstrating it to be a versatile toolchain for both attackers and defenders. Future work may extend its capabilities to post-synthesis netlist functionalities and various hardware platforms, further enhancing its adaptability and effectiveness. Expanding LLM applications into the analog domain, Chaudhuri et al. proposed LATENT~\cite{chaudhuri2025latentllmaugmentedtrojaninsertion}, a feedback-guided agentic framework that autonomously inserts stealthy analog hardware Trojans into A/MS netlists using a Thought-Action-Observation loop and SPICE-based simulation feedback. LATENT achieves low activation ranges (15.7\%) and area overhead (7.4\%) while inducing significant output degradation (11.3\%), outperforming fixed-pattern Trojans like A2 and DELTA. However, extending LATENT to support AC and small-signal behavior could improve its applicability to a broader range of analog threat scenarios.


Generative Trojan design also became more automated in 2025. Faruque et al. introduced GHOST (Generator for Hardware-Oriented Stealthy Trojans), which orchestrates hardware-Trojan generation and insertion entirely through systematic prompt engineering---combining role-based, contextual, and reflexive-validation prompting---and shows that GPT-4 can successfully insert Trojans in 88.9\% of attempts across RTL designs, frequently evading structural machine-learning detectors~\cite{faruque2025unleashing}. On the defensive side, Thorat et al. proposed TROJAN-GUARD, a graph-neural-network detector that generates embeddings for large RTL designs with model quantization for efficient training and inference, reaching 98.66\% precision and 92.30\% recall and directly addressing the poor scaling of earlier GNN-based detectors~\cite{thorat2025trojanguard}. Read alongside LATENT's agentic analog-Trojan insertion~\cite{chaudhuri2025latentllmaugmentedtrojaninsertion} and SPICED+'s LLM-enhanced analog-Trojan correction~\cite{chaudhuri2025spiced}, these works underscore the dual-use nature of LLMs in the Trojan landscape, where the same generative and reasoning capabilities serve both attackers and defenders.


\subsubsection{Logic Obfuscation}
Researchers have explored leveraging LLMs for logic obfuscation, with one notable work by Latibari et al.introducing Obfus-chat, an innovative framework utilizing GPT models to automate the obfuscation of hardware IPs~\cite{10539877}. This system takes hardware design netlists and key sizes as inputs, producing obfuscated code to enhance security. The framework's effectiveness is assessed using the Trust-Hub Obfuscation Benchmark, incorporating SAT attacks and functional verification to ensure security and design integrity.

\subsubsection{Side-Channel Attack}
Side-channel attacks (SCAs) remain a critical threat to hardware security, with power and cache-based attacks being among the most prominent.
Researchers have leveraged LLMs and graph-based techniques to address side-channel vulnerabilities, particularly power and cache-based attacks. These approaches automate vulnerability detection, generate protective measures, and improve the interpretability of security analyses, enabling early-stage vulnerability identification and reducing the need for costly post-silicon fixes.
A significant focus has been on power side channel (PSC) attacks, which exploit power consumption patterns to leak sensitive information.  Srivastava et al. addressed this challenge with SCAR, a pre-silicon PSC analysis framework that utilizes GNNs to convert designs into control-data flow graphs for detecting modules susceptible to side-channel leakage~\cite{10508974}. The framework includes a deep-learning-based explainer for providing human-accessible explanations of detection decisions and a fortification component that uses LLMs to generate and insert protective design code at vulnerable points. SCAR achieves significant figures of 94.49\% localization accuracy, 100\% precision, and 90.48\% recall on various encryption algorithms, including AES, RSA, PRESENT, Saber, and CRYSTALS-Kyber. This significantly improves design robustness and efficiency by enabling early vulnerability detection and reducing the need for costly post-silicon fixes. Additionally, SCAR reduces features for GNN model training by 57\% while maintaining comparable accuracy.

Cache-based side-channel attacks have also been a focal point, with researchers leveraging LLMs to model and evaluate secure cache architectures. He et al. proposed a novel probabilistic information flow graph to model interactions between victim and attacker programs and the cache architecture~\cite{he2017secure}. It introduces the Probability of Attack Success (PAS) metric to quantitatively assess a cache's resilience against various cache side-channel attacks. This model and metric are applied to evaluate nine different cache architectures across four classes of such attacks. LLMs are leveraged to automate the analysis and generation of potential attack scenarios. This approach enables verification and comparison of different secure cache architectures' resilience to side-channel attacks without requiring simulation or hardware implementation.

Table ~\ref{tab:llm_based_secure_hardware_design} summarizes the latest contributions that leverage LLMs for secure hardware design. This table lists more than two dozen influential works from 2023 and 2025, showing how researchers apply LLMs to various security applications, including vulnerability detection, SoC security, Trojan insertion, logic obfuscation, and side-channel analysis. The works utilize advanced AI models such as GPT-3.5 Turbo, GPT-4, HS-BERT, code-davinci-002, Llama-70B, and GNNs. Each paper showcases the ability of these models to tackle specific security challenges, enhancing hardware resilience and protecting against emerging threats. These contributions highlight how researchers increasingly rely on LLMs and attention-based methods to innovate and strengthen hardware security practices.

\begin{table*}[!t]
\centering
\caption{LLM-Based Secure Hardware Design}
\label{tab:llm_based_secure_hardware_design}
\scriptsize
\renewcommand{\arraystretch}{1.15}
\setlength{\tabcolsep}{3pt}

\begin{tabularx}{\textwidth}{|
>{\centering\arraybackslash}m{1cm}|
>{\raggedright\arraybackslash}p{3.6cm}|
>{\centering\arraybackslash}m{1.3cm}|
>{\centering\arraybackslash}m{1.3cm}|
>{\centering\arraybackslash}m{1.3cm}|
>{\centering\arraybackslash}m{1.3cm}|
>{\centering\arraybackslash}m{1.3cm}|
>{\raggedright\arraybackslash}X|}
\hline

\rowcolor[HTML]{E3E3E3}
\textbf{Year} &
\textbf{Method Name} &
\textbf{Vulnerability Detection} &
\textbf{SoC Security} &
\textbf{Trojan Insertion} &
\textbf{Logic Obfuscation} &
\textbf{Side Channel} &
\textbf{AI Model} \\
\hline

2023 & Thakur et al.~\cite{thakur2023verigen} & & \checkmark & & & & GPT-3.5-Turbo \\ \hline
2023 & Saha et al.~\cite{saha2023llm} & \checkmark & & & & & GPT-3 and GPT-4 \\ \hline
2023 & Meng et al.~\cite{meng2023unlocking} & & \checkmark & & & & HS-BERT \\ \hline
2023 & Kande et al.~\cite{kande2023llmassisted} & \checkmark & & & & & code-davinci-002 \\ \hline

2024 & Saha et al. (Vul-FSM)~\cite{10545393} & \checkmark & \checkmark & & & & GPT-3.5-Turbo \\ \hline
2024 & Gadde et al.~\cite{gadde2024artificialintelligencegenailens} & \checkmark & & & & & GPT-3.5-Turbo \\ \hline
2024 & Akyash et al.~\cite{akyash2024selfhwdebug} & \checkmark & \checkmark & & & & Llama3-70B \\ \hline
2024 & Tarek et al.~\cite{tarek2024socurellm} & & \checkmark & & & & GPT-4 \\ \hline
2024 & Kardakis et al.~\cite{kokolakis2024harnessing} & & \checkmark & \checkmark & & & GPT-3.5/4 \\ \hline
2024 & Hayashi et al.~\cite{hayashi2024hardware} & & & \checkmark & & & GPT-4 \\ \hline
2024 & Latibari et al.~\cite{10539877} & & & & \checkmark & & GPT-3.5 \\ \hline
2024 & Bhandari et al.~\cite{bhandari2024sentaursecurityenhancedtrojan} & & & \checkmark & \checkmark & & GPT-4 \\ \hline
2024 & Srivastava et al.~\cite{10508974} & & \checkmark & & & \checkmark & GNN \\ \hline

2025 & Ghimire et al.~\cite{Ghimire2025_Hwrex} & \checkmark & & & & & GPT-3.5-Turbo \\ \hline
2025 & Chaudhuri et al. (LATENT)~\cite{chaudhuri2025latentllmaugmentedtrojaninsertion} & & & \checkmark & & & GPT-4o-mini \\ \hline
2025 & Tarek et al.~\cite{tarek2025bugwhisperer} & & \checkmark & & & & LLMs \\ \hline
2025 & Faruque et al. (Ghost)~\cite{faruque2025unleashing} & & & \checkmark & & & LLMs \\ \hline
2025 & Chaudhuri et al. (Spiced+)~\cite{chaudhuri2025spiced} & & \checkmark & & \checkmark & & LLMs \\ \hline
2025 & MARVEL~\cite{collini2025marvel} & \checkmark & \checkmark & & & & LLM multi-agent \\ \hline
2025 & LASHED~\cite{ahmad2025lashed} & \checkmark & & & & & LLMs + static analysis \\ \hline
2025 & VerilogLAVD~\cite{long2025veriloglavd} & \checkmark & & & & & LLMs, VeriPG rules \\ \hline
2025 & LASA~\cite{ankireddy2025lasa} & \checkmark & \checkmark & & & & LLM + RAG + FPV \\ \hline
2025 & SV-LLM~\cite{saha2025svllm} & \checkmark & \checkmark & & & & LLM multi-agent \\ \hline
2025 & TROJAN-GUARD~\cite{thorat2025trojanguard} & & & \checkmark & & & GNN, quantized \\ \hline
2025 & MALLS~\cite{talukdar2025malls} & \checkmark & & \checkmark & & & LLM \\ \hline

\end{tabularx}
\end{table*}

Across the hardware-security literature, the strongest results tend to emerge when LLMs are coupled with structured security artifacts rather than used in isolation. In vulnerability detection and policy extraction, domain-specific datasets and security-aware prompts improve the model’s ability to reason about hardware-specific threat patterns. In verification and debugging, LLM outputs become substantially more reliable when checked against formal properties, simulation feedback, or curated benchmark suites. In Trojan analysis and side-channel mitigation, graph structure, netlist semantics, and external validation remain critical because security-relevant behavior is often not fully visible in plain text alone. Overall, these studies suggest that LLMs are most effective in hardware security when they serve as reasoning and coordination layers over formal tools, structured datasets, and domain-specific representations, rather than as purely stand-alone generators~\cite{bhandari2024sentaursecurityenhancedtrojan},\cite{gadde2024artificialintelligencegenailens},\cite{kande2023llmassisted},\cite{meng2023unlocking},\cite{10508974},\cite{tarek2024socurellm}.

\subsection{Other Attention-based Hardware Security }

Attention-based models have played a pivotal role in transforming hardware security, enabling more advanced threat detection and mitigation strategies.  

A common theme across recent works is the use of graph-based representations. For instance, Chen et al. proposed TrojanFormer, a resource-efficient method for hardware Trojan detection using a graph transformer network. It converts HDL designs into graph structures to enhance accuracy and efficiency and employs a NodeFormer-based network with linear-complexity message passing and edge-regularization loss. TrojanFormer was evaluated on a Trust-Hub dataset with over 100,000 nodes, achieving a 97.66\% F1 score on small and medium ICs. For large ICs, it showed a 4\% accuracy improvement and an 18\% reduction in computational overhead. Key limitations include difficulty capturing long-range dependencies in large ICs and reduced performance on circuits with sparse or incomplete graph topologies~\cite{10672762}.
Another key direction is the integration of transformer models with graph-based pre-processing layers.
For instance, Li e t al. proposed a transformer-based method that uses both a transformer and a Graph Convolutional Network as a pre-processing layer for detecting and localizing HTs at the RTL level and tested their architecture on the TrustHub dataset. They use RTL features to detect and localize HT attacks~\cite{10317971}.
In another endeavor, Zhang et al. proposed a novel bit-level Hardware Trojan localization approach, namely B-HTRecognizer, which leveraged Graph Attention Networks. It converted the RTL Verilog code into bit-level edge-featured Data Flow Graphs to capture multi-dimensional feature representation, thus enhancing the precision of HT detection and localization. They presented TrustHub IMEex, a scaled open-source dataset for ML training. B-HTRecognizer achieved 84\% precision, 93\% recall for non-cross-designs, and 77\% recall for cross-design tests on RISC-V, proving effective for various designs. It characterized the progress in automation and fine-grained localization for pre-silicon hardware security~\cite{10802937}.
Continuing the graph-attention theme, Yasaei et al. proposed a hardware-Trojan detection method that integrates the GraphSAGE algorithm with a graph-Transformer structure, embedding a multi-head attention mechanism over gate-level-netlist graphs in which nodes represent logic gates and edges represent interconnections~\cite{yasaei2025hardware}. Evaluated on Trust-Hub benchmarks, the model reaches a 96.60\% true-positive rate, a 99.99\% true-negative rate, and a 97.66\% F1 score, illustrating how combining neighborhood aggregation with global attention strengthens structural Trojan detection beyond conventional GNNs.

Finally, we note that few works have considered evaluating the quality of LLM-generated hardware designs from the security perspective, with recent work by Afsharmazayejani et al. providing an initial exploration of how such benchmarking could be undertaken~\cite{10691745}.

\section{Discussion and Future Directions}
\label{sec:discussion}
At the current stage of the field, attention-based models are most mature in tasks that provide strong intermediate feedback and well-structured outputs. In digital hardware design, this includes specification-to-RTL generation for relatively bounded modules, iterative repair of HDL code, assertion and testbench generation, and debugging assistance supported by compiler, simulator, or formal feedback~\cite{gao2024autovcodersystematicframeworkautomated,nakkab2024romebuiltsinglestep,qiu2024autobenchautomatictestbenchgeneration,thorat2023advanced,10.1145/3649329.3657353,yao2024hdldebugger}. In non-LLM attention-based hardware design, the strongest results appear in graph-structured tasks such as placement, floorplanning, hotspot detection, logic synthesis, and circuit representation learning, where relational structure is explicit and optimization objectives are well defined~\cite{deng2024morehopwisegraphattention,lai2023chipformertransferablechipplacement,li2024circuittransformerendtoendcircuit,yang2024floorplanning,9643590}. In hardware security, the most compelling applications currently include vulnerability detection, security-property extraction, side-channel analysis, and structured verification tasks in which LLMs or graph-based models can be coupled with formal checks, datasets, or security-specific representations~\cite{gadde2024artificialintelligencegenailens,kande2023llmassisted,meng2023unlocking,10508974,tarek2024socurellm}.

At the same time, several important capabilities remain promising but not yet mature. Recent work on multimodal Verilog generation, architecture-specification assistance, EDA workflow orchestration, and analog-design optimization suggests that attention-based systems may eventually become strong coordination layers across specification, implementation, verification, and layout~\cite{chang2024naturallanguageenoughbenchmarking,he2024chatedalargelanguagemodel,jiang2024iicpilotintelligentintegratedcircuit,li2024specllm,liu2024layoutcopilotllmpoweredmultiagentcollaborative,yin2024adollmanalogdesignbayesian}. However, these capabilities remain uneven. Many systems still depend on carefully structured prompts, narrow benchmarks, domain-specific datasets, or substantial human intervention when task complexity increases~\cite{chang2024naturallanguageenoughbenchmarking,he2024chatedalargelanguagemodel,jiang2024iicpilotintelligentintegratedcircuit,yang2023new}. Likewise, end-to-end one-shot chip design without strong external feedback remains unrealistic, especially for complex multi-module systems, analog flows, or security-critical designs~\cite{10691811,chang2024naturallanguageenoughbenchmarking,yang2023new}. For this reason, attention-based systems should not yet be viewed as replacements for formal verification, simulation, or specialized security analysis; instead, they are most reliable when used as reasoning and orchestration layers over those tools~\cite{gadde2024artificialintelligencegenailens,kande2023llmassisted,qiu2024autobenchautomatictestbenchgeneration}.

A major future direction for the field is the transition from sequential design and security evaluation to true security-aware hardware design. Today, many design systems optimize for power, performance, area, or functional correctness first, while security checks are introduced later as separate verification or analysis steps~\cite{he2024chatedalargelanguagemodel,jiang2024iicpilotintelligentintegratedcircuit,thorat2023advanced}. Conversely, many hardware-security works assume the design artifact is already available and then focus on vulnerability detection, policy extraction, Trojan analysis, or side-channel mitigation~\cite{gadde2024artificialintelligencegenailens,meng2023unlocking,10508974,tarek2024socurellm}. A more transformative direction is to treat security constraints, performance objectives, and verification requirements as coequal optimization targets throughout the design loop. In such a workflow, generated RTL would be evaluated not only for functionality and PPA, but also for attack surface, formal security properties, side-channel risk, and trustworthiness under realistic deployment conditions.

Another transformative direction is the development of AI-native closed-loop EDA systems. Rather than using LLMs only as front-end code generators, future systems should be built around persistent interaction with simulators, formal engines, synthesis tools, layout tools, and security analyzers. In this model, the attention-based system becomes a reasoning and orchestration layer over the design flow, not just a text generator. Such systems would continuously revise candidate designs in response to implementation feedback, proof failures, timing violations, security-property violations, and side-channel indicators. If successful, this would move the field from prompt-based assistance to genuinely adaptive design automation~\cite{he2024chatedalargelanguagemodel,jiang2024iicpilotintelligentintegratedcircuit,li2024circuittransformerendtoendcircuit,qiu2024autobenchautomatictestbenchgeneration,10508974,thorat2023advanced}.

A third important direction is secure-by-construction hardware generation. Current literature shows that LLM-generated hardware can still contain CWEs, weak security properties, or vulnerabilities that only become visible under formal analysis~\cite{gadde2024artificialintelligencegenailens,kande2023llmassisted}. A more ambitious goal is to generate hardware together with security invariants, assertions, policy constraints, and attack-surface estimates from the outset, rather than appending these checks later. This would turn hardware security from a retrospective validation step into a generative design primitive. Progress in SoC documentation analysis, assertion generation, and structured vulnerability detection suggests that the ingredients for this transition are emerging, but they are not yet integrated into a unified secure-design methodology~\cite{gadde2024artificialintelligencegenailens,kande2023llmassisted,meng2023unlocking,saha2023llm,tarek2024socurellm}.

Finally, the field will need stronger multimodal circuit models and more trustworthy deployment mechanisms. Hardware design and security are inherently multimodal, spanning natural-language specifications, HDL, netlists, waveforms, simulation traces, layout views, and security documentation. Future models that align these modalities could reason more effectively across abstraction levels than text-only systems~\cite{chang2024naturallanguageenoughbenchmarking,liu2024layoutcopilotllmpoweredmultiagentcollaborative}. However, greater capability must be paired with stronger safeguards, benchmark quality, constrained tool permissions, interpretable failure analysis, and clear human-in-the-loop oversight rather than unrestricted autonomy.


A recent frontier-model case further underscores why trustworthy deployment and staged access should become central design principles for AI-assisted hardware workflows. In April 2026, Anthropic announced Claude Mythos Preview as its most capable model to date for cybersecurity-related tasks, but did not deploy it for general access; instead, the model was released only through limited research and defensive-security programs such as Project Glasswing. Anthropic’s public risk materials state that Mythos Preview was first deployed internally and then released to only a small set of external customers through limited research access, explicitly noting that it has not been deployed for general access. Independent evaluation by the UK AI Security Institute also reported that Mythos Preview showed substantial gains in offensive cyber capability, including strong expert-level capture-the-flag performance and the ability to complete multi-step attack simulations that earlier frontier models could not reliably solve. Although this example is not specific to hardware design, it is directly relevant to hardware-security research because it demonstrates that as frontier models become more capable in cyber tasks, broader deployment decisions increasingly require safeguards, constrained access, and rigorous validation rather than capability alone~\cite{anthropic2026mythosrisk,anthropic2026glasswing,aisi2026mythoseval}.

\section{Conclusion}
\label{sec:conclusion}
Integrating attention-based models, especially LLMs, into hardware design and security is at a transformative stage, offering groundbreaking automation, efficiency, and robustness capabilities. This paper has presented a detailed investigation into how these models are applied to automate HDL generation, optimize chip design workflows, and enhance security mechanisms against threats such as hardware Trojans and side-channel attacks. The state-of-the-art survey shows the tremendous improvement made so far and the challenges ahead, such as data scarcity and the limitation of general-purpose models applied to hardware-specific tasks. Although the latest achievements in attention mechanisms and fine-tuned models have provided significant possibilities, challenges persist, including the need for domain-specific datasets and lightweight models. 

\bibliographystyle{IEEEtran}
\bibliography{LLMSurvey}

\section{Biography Section}

\vspace{-6em}
\begin{IEEEbiography}
[{\includegraphics[width=1in,height=1.25in,clip,keepaspectratio]{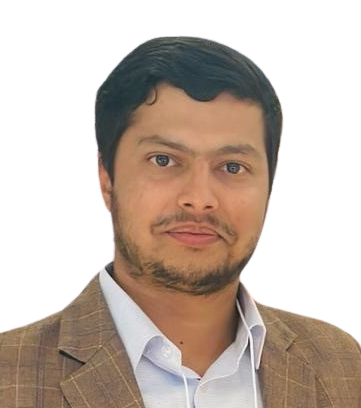}}]{Sujan Ghimire}received the B.E. degree in Computer Engineering from Tribhuvan University, Nepal, in 2019 and the M.S. degree in Software Engineering from The University of Arizona, Tucson, AZ, USA, in 2026. He is currently pursuing the Ph.D. degree in Software Engineering with the Department of Electrical and Computer Engineering, The University of Arizona. His research interests include large language models for hardware design and security, hardware and system security, binary analysis and hardening, trustworthy artificial intelligence, and Internet of Things (IoT) security.
\end{IEEEbiography}

\vspace{-2em}
\begin{IEEEbiography}
[{\includegraphics[width=1in,height=1.25in,clip,keepaspectratio]{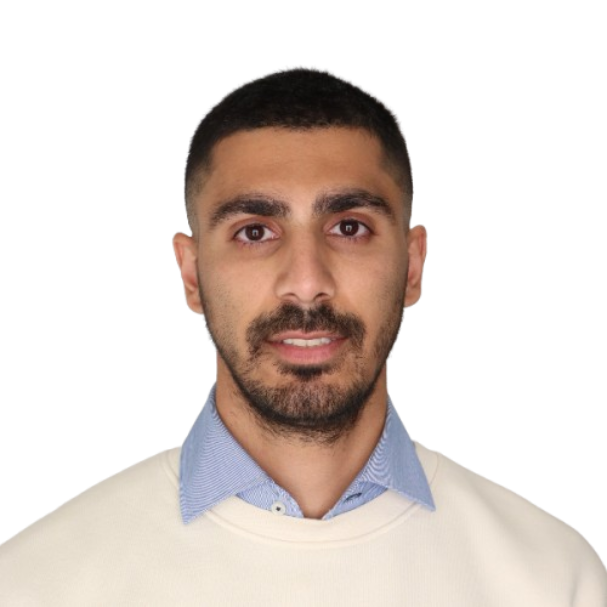}}]{Parsa Mirfasihi,} received the B.S. degree in Electrical and Computer Engineering from Iran University of Science and Technology, Tehran, Iran, in 2022 and the M.S. degree in Electrical and Computer Engineering from San Francisco State University, San Francisco, CA, USA, in 2026. He is currently pursuing the Ph.D. degree in Electrical and Computer Engineering at the University of Arizona, Tucson, AZ, USA. His research interests include hardware design and security, trustworthy artificial intelligence, transformer-based architectures, and AI-driven approaches for cybersecurity and hardware security.
\end{IEEEbiography}

 \vspace{-4em}
\begin{IEEEbiography}
[{\includegraphics[width=1in,height=1.25in,clip,keepaspectratio]{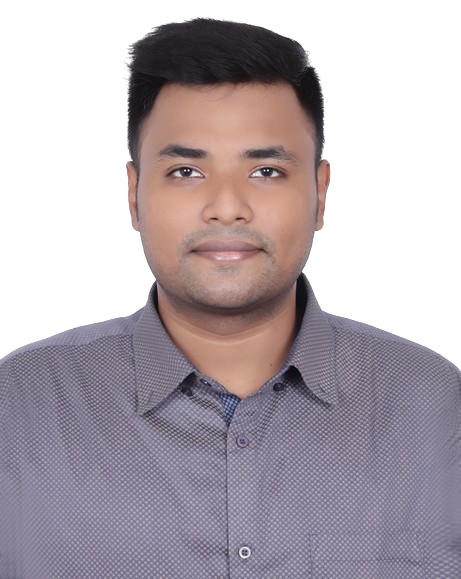}}]{Muhtasim Alam Chowdhury} received the B.S. degree in electrical and
electronics engineering from North South University in 2019. He is currently
pursuing a Ph.D. degree with the PRISM Laboratory; Electrical and Computer Engineering
Department, The University of Arizona. His research interests are IoT
hardware supply chain security, hardware security, in-memory computing,
spin-based devices, AI applications for security, and robotics.
\end{IEEEbiography}

 \vspace{-4em}
\begin{IEEEbiography}
[{\includegraphics[width=1in,height=1.25in,clip,keepaspectratio]{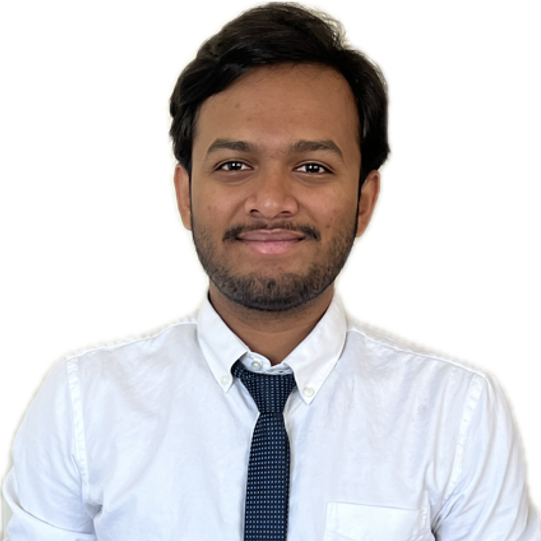}}]{Harish Kumar Dharavath} received the B.Tech. degree in Electrical and Computer Engineering from KL University, India, in 2022, and the M.S. degree in ECE from Arizona State University in 2024. He is currently pursuing thhe Ph.D. degree with the PRISM Laboratory, Electrical and Computer Engineering Department, The University of Arizona. His research interests are VLSI Design, hardware security, LLMs for Chip Design and security, Transformer architecture security, and Photonic computing.
\end{IEEEbiography}

 \vspace{-4em}
\begin{IEEEbiography}[{\includegraphics[width=1.2in, height=1.3in, clip, keepaspectratio]{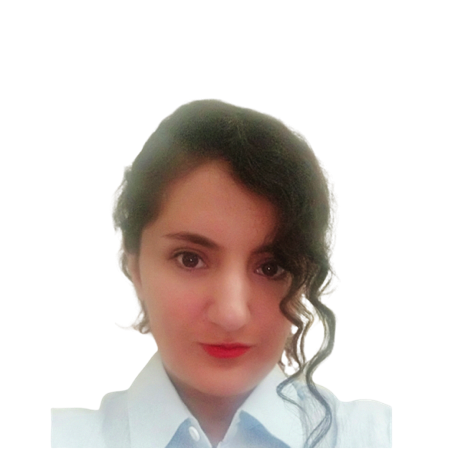}}]{Banafsheh Saber Latibari} received a B.Sc. degree in Computer Engineering from the K. N. Toosi University of Technology and the M.Sc. degree in Computer Architecture from the Sharif University of Technology. She received her Ph.D. degree from the Electrical and Computer Engineering (ECE) Department at the University of California, Davis. She is currently a Postdoctoral Research Associate at the University of Arizona's ECE Department. Her research focuses on deep learning, embedded system security, and computer architecture.
    
\end{IEEEbiography}

\vspace{-4em}
\begin{IEEEbiography}[{\includegraphics[width=1.1in, height=1.3in, clip, keepaspectratio]{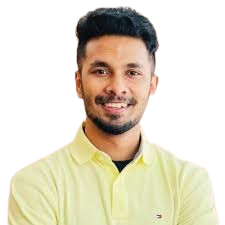}}]{Muntasir Mamun} is pursuing his Ph.D. degree in Systems and Industrial Engineering department in Software Engineering Program. He completed his M.Sc in Computer Science from the University of South Dakota in 2022 and a B.Sc in Electrical and Electronic Engineering from American International University-Bangladesh in 2018. His research primarily focused on but not limited to cybersecurity, ML, anomaly behavior analysis, healthcare, and generative AI.
\end{IEEEbiography}

\vspace{-4em}
\begin{IEEEbiography}[{\includegraphics[width=0.9in, height=1in,clip, keepaspectratio]{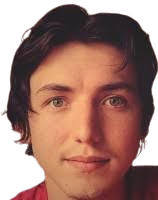}}]{Jaeden Carpenter} received the B.S. degree in Electrical and Computer Engineering from The University of Arizona, Tucson, AZ, USA, in 2024 and the M.S. degree in Electrical and Computer Engineering from The University of Arizona in 2025. He is currently a Firmware Engineer with IBM. Previously, he served as a Graduate Teaching Assistant and a Semiconductor Researcher with the PRISM Laboratory at The University of Arizona. His research interests include semiconductor education, semiconductor design, firmware development, and hardware systems.
\end{IEEEbiography}

\vspace{-4em}
\begin{IEEEbiography}[{\includegraphics[width=1in, height=1.25in, clip, keepaspectratio]{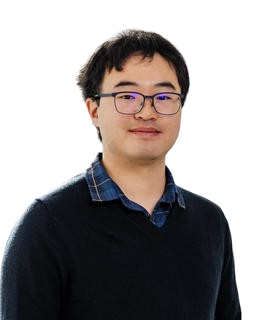}}]{Benjamin Tan}
received the B.E. (Hons.) degree in computer systems engineering and the Ph.D. degree from the University of Auckland, Auckland, New Zealand, in 2014 and 2019, respectively. He was a Professional Teaching Fellow with the Department of Electrical and Computer Engineering, University of Auckland, in 2018. From 2019 to 2021 he was with New York University, Brooklyn, NY, USA, where he was a Postdoctoral Associate and then a Research Assistant Professor affiliated with the NYU Center for Cybersecurity. He is currently an Assistant Professor at the University of Calgary, Calgary, AB, Canada. His research interests include computer engineering, hardware security, and electronic design automation. Dr. Tan is a Professional Member of the Association of Professional Engineers and Geoscientists of Alberta (APEGA). He is a member of IEEE and ACM. 
 \end{IEEEbiography}
 
\vspace{-4em}

\begin{IEEEbiography}[{\includegraphics[width=1in, height=1.25in, clip, keepaspectratio]{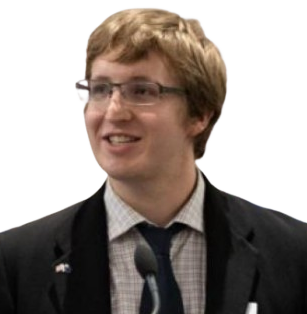}}]{Hammond Pearce} is a Lecturer (Assistant Professor) at UNSW Sydney's School of Computer Science and Engineering. His main research focus is in the cybersecurity of embedded and cyber-physical systems (CPS), including in additive manufacturing and in industrial informatics, and also examines the implications of ML on design and applications in this space. He received a B.E. (Hons) degree in Computer Systems Engineering in 2014 and a Ph.D. in Computer Systems Engineering in 2020, both from the University of Auckland, Auckland, New Zealand. In 2019, he took part in the NASA International Internship Programme and worked at NASA Ames in California, and he has also worked in several industry positions, including as a full-stack web developer and as an electronics contractor working on Li-ion battery management systems.  Previously, he has also worked as a Research Assistant Professor at NYU Tandon's Department of Electrical and Computer Engineering and in the NYU Center for Cybersecurity. His other research interests include IoT, CPS, compilers, and AI / ML.
\end{IEEEbiography}

\vspace{-4em}
\begin{IEEEbiography}[{\includegraphics[width=1in, height=1.25in, clip, keepaspectratio]{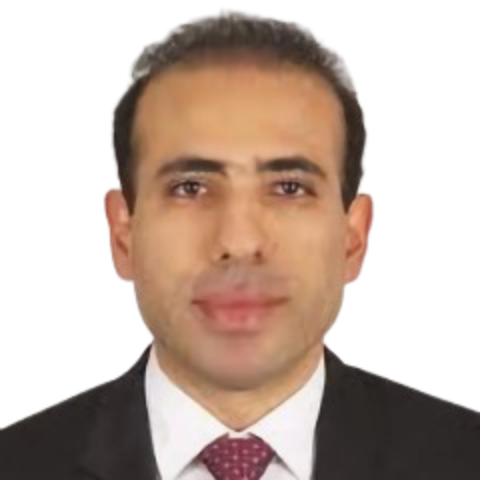}}]{Farshad Firouzi} is a Research Associate Professor with the School of Electrical, Computer and Energy Engineering, Arizona State University, Tempe, AZ, USA. Prior to joining Arizona State University, he held research positions with KU Leuven, IMEC, Karlsruhe Institute of Technology (KIT), Semiconductor Research Corporation (SRC), and SI LAB. His research interests include machine learning, neuromorphic computing, Internet of Things (IoT), hardware security, big data analytics, and intelligent electronic systems. He has authored more than 40 journal and conference publications and has served the research community in various editorial and conference leadership roles, including reviewer, guest editor, program chair, and technical program committee member for numerous journals and conferences.
\end{IEEEbiography}

\vspace{-3em}
 \begin{IEEEbiography}[{\includegraphics[width=1in, height=1.25in, clip, keepaspectratio]{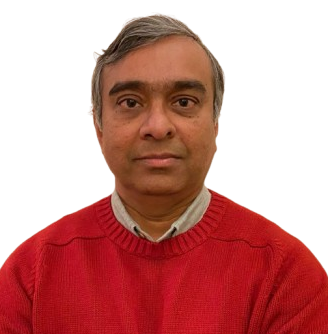}}]{Krishnendu Chakrabarty}(Fellow, IEEE) is the Fulton Professor of Microelectronics in the School of Electrical, Computer and Energy Engineering at Arizona State University (ASU), where he also serves as the Chief Technology Officer (CTO) of the Department of Defense Microelectronics Commons Southwest Advanced Prototyping (SWAP) Hub and the Director of the ASU Center on Semiconductor Microelectronics. He received his B.Tech. degree from the Indian Institute of Technology Kharagpur in 1990 and his M.S.E. and Ph.D. degrees from the University of Michigan, Ann Arbor, in 1992 and 1995, respectively. Dr. Chakrabarty is widely recognized for his contributions to the field of microelectronics, with research interests spanning integrated circuit (IC) testing, microfluidics-based biochips, and hardware security. He is a Fellow of ACM and the American Association for the Advancement of Science (AAAS), and has also been honored as a Golden Core Member of the IEEE Computer Society.
\end{IEEEbiography}

\vspace{-2em}
\begin{IEEEbiography}[{\includegraphics[width=1in, height=1.25in, clip, keepaspectratio]{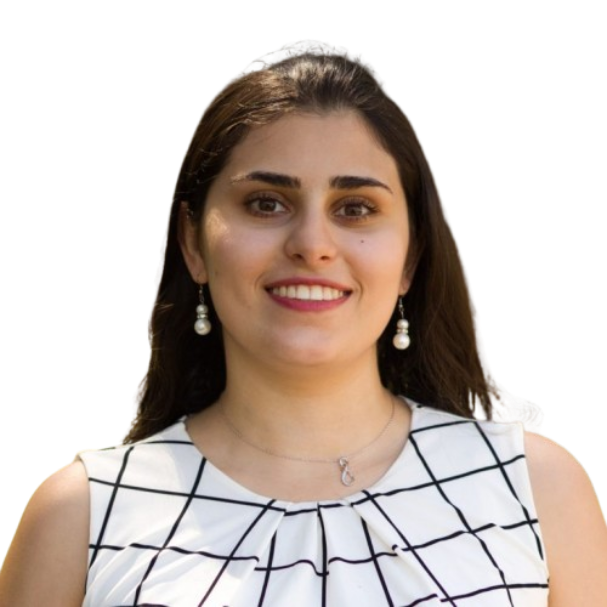}}]{Rozhin Yasaei} received the B.S. degree in Electrical Engineering from Sharif University of Technology, Tehran, Iran, in 2018, the M.S. degree in Computer Engineering from the University of California, Irvine, CA, USA, in 2021, and the Ph.D. degree in Computer Engineering from the University of California, Irvine, in 2023. She is currently an Assistant Professor with the Cyber Operations, Intelligence, and Technology Program, The University of Arizona, Tucson, AZ, USA. Her research interests include hardware security, embedded systems security, cyber-physical systems security, electronic design automation, and the application of machine learning and graph neural networks to secure and trustworthy computing systems.
\end{IEEEbiography}

\vspace{-2em}
\begin{IEEEbiography}[{\includegraphics[width=1in, height=1.25in, clip, keepaspectratio]{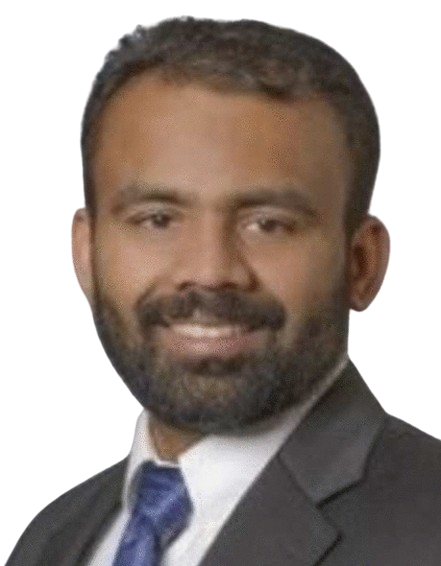}}]{Pratik Satam} is an Assistant Professor in the Department of Systems and Industrial Engineering at the University of Arizona, Tucson, AZ, USA. He is also part of the new Software Engineering program in College of Engineering at the University of Arizona. He received his B.E. degree in Electronics and Telecommunication Engineering from the University of Mumbai, in 2013, and his M.S. and Ph.D. degrees in Electrical and Computer Engineering from the University of Arizona in 2015 and 2019, respectively. From 2019 to 2022, he was a Research Assistant Professor at the Department of Electrical and Computer Engineering, at the University of Arizona. His current research interests include autonomic computing, cyber security, cyber resilience, secure critical infrastructures, and cloud security. He is an Associate Editor for the scientific journal Cluster Computing.
\end{IEEEbiography}

\begin{IEEEbiography}[{\includegraphics[width=1in, height=1.25in, clip, keepaspectratio]{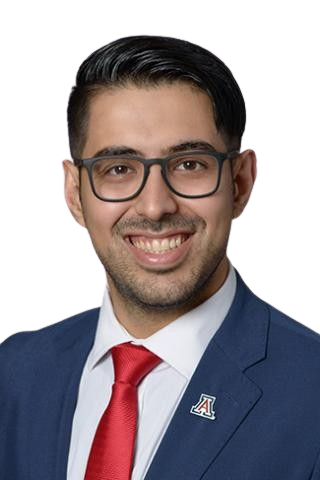}}]{Soheil Salehi} (Member, IEEE) is an assistant professor in the Electrical and Computer Engineering (ECE) Department and Fellow of Center for Semiconductor Manufacturing (CSM) at the University of Arizona (UofA). Prior to joining the UofA, Soheil was an NSF-Sponsored Computing Innovation Fellow in the Accelerated, Secure, and Energy-Efficient Computing Laboratory and the Center for Hardware and Embedded Systems Security and Trust at the University of California, Davis (UC Davis). He received his Ph.D. and M.S. degrees in ECE from the University of Central Florida (UCF) in 2016 and 2020, respectively. He has expertise in the areas of hardware security and IoT supply-chain security as well as applied ML for secure hardware design. Moreover, he has designed novel circuits and architectures for secure and accelerated computing. He has received several nominations and award recognition, which include the Outstanding Reviewer Award at IEEE/ACM DAC'23, the Best Poster Award at ACM GLSVLSI'19, the Best Paper Award Nominee at IEEE ISQED'17 as well as the Best Presentation at UC Davis Postdoctoral Research Symposium in 2021, and the Best Graduate Teaching Assistant Award at UCF in 2016. 
\end{IEEEbiography}

\newpage

\vfill

\end{document}